\documentclass[12pt]{article}

\usepackage[a4paper, margin=1in]{geometry}

\usepackage{amsmath, amssymb, amsfonts}

\usepackage{graphicx}
\usepackage{float}
\usepackage{subcaption}

\usepackage{setspace}   
\usepackage{lineno}     

\usepackage[
  colorlinks=true,
  linkcolor=blue,
  citecolor=blue,
  urlcolor=blue
]{hyperref}

\graphicspath{{./figures/}}

\newcommand{\figref}[1]{Fig.~\ref{fig:#1}}

\title{Limits of the Formal Integrals of Motion}

\author{
G. Contopoulos$^{1}$,
A. C. Tzemos$^{1}$,
F. Zanias$^{1,2}$\\[4pt]
$^{1}$Research Center for Astronomy and Applied Mathematics,\\
Academy of Athens, Soranou Efessiou 4, GR-11527 Athens, Greece\\
$^{2}$University of Amsterdam, Science Park 904,\\
1098 XH Amsterdam, The Netherlands\\
}
\date{} 

\begin{document}

\maketitle

\begingroup
\renewcommand{\thefootnote}{}
\footnote{Emails: gcontop@academyofathens.gr,
 atzemos@academyofathens.gr, foivos.zanias@student.uva.nl
}
\endgroup

\begin{abstract}
We consider a formal (approximate) integral of motion in Hamiltonians of the form $H=\frac{1}{2}(X^2+Y^2+\omega_1^2x^2+\omega_2^2y^2)+\epsilon(\eta xy^2+\alpha x^3+\beta x^2y+\gamma y^3)$ generalizing previous cases with $\beta=\gamma=0$. First we give the general form of this integral when  $\omega_1/\omega_2$ is irrational and then we consider the case of commensurable frequencies. In particular we study the integrals for the resonances $\omega_1/\omega_2=4/1,  5/1, 3/2, 4/3, 3/1$ and $2/1$. We also calculate the invariant curves and the orbits in the cases $\omega_1/\omega_2=2/1$ and $1/1$ (with $\beta=\gamma=0$) and we compare  the exact-numerical and the theoretical results predicted by the formal integral when $\beta\gamma\neq0$. In the special  case $\omega_1/\omega_2=1/1$ we find an integral when $\beta=\gamma=0$ and $\eta\alpha\neq0$ or $\eta=\alpha=0$ and $\beta\gamma\neq 0$, but this is not possible when $\eta\alpha\beta\gamma\neq 0$.  However, we find that the invariant curves and the orbits can be approximated by a non-resonant integral with $\omega_1/\omega_2=5\sqrt{2}/7=1.010\dots$.
\end{abstract}

\section{Introduction}\label{sec1}
In a Hamiltonian dynamical system, the phase space is typically a mixed structure, consisting of regions with initial conditions giving rise to organized trajectories and others leading to chaotic trajectories. 

The concept of integrability in dynamical systems has been a subject of extensive study and remains an active field of research. A Hamiltonian system with $N$ degrees of freedom is considered integrable if it possesses $N$ independent integrals of motion that are in involution, meaning that their Poisson brackets vanish pairwise. As a consequence, the equations of motion allow for separation of variables, making it possible to express their solutions in terms of analytic integrals. The phase space of an integrable system is foliated by $N$-dimensional invariant tori, on which the motion is quasi-periodic. Systems with more than $N$ integrals of motion are called superintegrable. Several cases of integrable and superintegrable systems have been studied in the past \cite{stackel1890,stackel1893,hietarinta1987direct,mitsopoulos2022quadratic,mitsopoulos2023cubic}. The breakdown of integrability is associated with the onset of chaotic behavior.

However, even in systems that deviate from strict integrability due to perturbations, one can often construct formal (approximate) integrals of motion in the form of non-convergent asymptotic series  \cite{Whittaker1916,Cherry1924a,Whittaker1937}. These formal integrals provide valuable insights into the system’s long-term behavior and the mechanisms that drive the transition from regular to chaotic  motion. In fact, formal integrals have been extensively studied and utilized in the field of Galactic Dynamics, particularly in the context of the ``third integral of motion''\cite{Contopoulos1960,lynden1962stellar}. This concept arises in the study of stellar orbits in non-axisymmetric potentials, where, in addition to energy and angular momentum (which are exact integrals in axisymmetric systems), a third approximately conserved quantity can often be identified when the perturbation is small. The existence of such a formal integral helps describe the structure of phase space and provides an understanding of the ordered components of stellar motion in galaxies \cite{contopoulos2002order}.

For a weak perturbation, the formal series approximately follow the exact orbits around a stable periodic orbit, provided they are truncated at a sufficiently high order. However, when the perturbation is strong, the theoretical orbits deviate from the exact ones, introducing chaos. According to the Nekhoroshev theorem \cite{nekhoroshev1977exponential}, there is an upper time limit of the number of terms of the formal integral beyond which the formal integrals of motion are no longer valid, and as a result, the corresponding truncated series yield worse rather than better results. Remarkably, the series of the formal integrals of motion are convergent close to unstable periodic orbits, where chaos emerges, as it was shown by Moser \cite{Moser1956,Moser1958} and Giorgilli \cite{Giorgilli2001}.

If we consider a 2D Hamiltonian of the form
\begin{equation}
    H=H_0+\sum_{n=1} \epsilon^n H_n=\frac{1}{2}(X^2+Y^2+\omega_x^2x^2+\omega_y^2y^2)+\epsilon H_1+\epsilon^2 H_2+\dots\label{genham}
\end{equation}
{with $H_n$ containing terms of degree $n+2$ in $x,y,X,Y$, (where $x,y$ are  the positions and $X,Y$ the corresponding momenta). Then the  formal }integral
\begin{equation}\label{phit}
    \Phi=\sum_{n=0}\epsilon^n\Phi_n
\end{equation}
can be constructed in an algorithmic way when the ratio $\omega_1 / \omega_2$ is irrational.{ Here, $\omega_1$ and $\omega_2$ denote the unperturbed frequencies associated with harmonic oscillations in the $x$- and $y$-directions, respectively, as defined by the quadratic part $H_0$ of the Hamiltonian. Their ratio determines whether the motion is \emph{non-resonant} or \emph{resonant}: if $\omega_1 / \omega_2$ is irrational, the two degrees of freedom oscillate quasiperiodically, and the system evolves on invariant tori. If $\omega_1 / \omega_2$ is rational, the system is said to be in resonance.}

{In the perturbative construction of the formal integral $\Phi$, each successive term involves solving linear equations that include denominators of the form $m \omega_1 \pm n \omega_2$, where $m, n \in \mathbb{Z}$. When the ratio $\omega_1 / \omega_2$ is irrational, these divisors never vanish for nonzero integers $m, n$, and the recursive procedure is well-defined. However, if $\omega_1 / \omega_2 = n/m$ is rational, some of these combinations vanish, and the corresponding terms in the expansion become divergent or introduce linear-in-time contributions, known as \emph{secular terms}. These terms violate the requirement that $\Phi$ be a constant of motion, and therefore the perturbative construction of a formal integral breaks down \cite{arnol2013mathematical}}.


In the rational cases the unperturbed Hamiltonian $H_0$ has an extra integral besides the integrals  
\begin{equation}
    \Phi_{1,0} = \frac{1}{2} X^2 + \omega_1^2 x^2, \quad \Phi_{2,0} = \frac{1}{2} Y^2 + \omega_2^2 y^2
\end{equation}
which is 
\begin{align}\label{eqdipli}
    \substack{S_0 \\ C_0}&=(2\Phi_{1,0})^{m/2}(2\Phi_{2,0})^{n/2}\substack{\sin \\ \cos}(n\omega_2t-m\omega_1(t- t_0))=(2\Phi_{1,0})^{m/2}(2\Phi_{2,0})^{n/2}\substack{\sin \\ \cos}(m t_0),
\end{align}  where $t_0$ is an initial time \footnote{{In Eq.~\eqref{eqdipli}, the notation 
\(\substack{S_0 \\ C_0} = \ldots \substack{\sin \\ \cos}(\ldots)\) 
means that both the sine and cosine expressions define conserved quantities.}}.  This is of degree $m+n$ in $x,y,X,Y$. Then we try  to construct a resonant form of the third integral by a proper combination of  $\Phi$ and this extra zero-order integral \cite{contopoulos2002order,ContopMouts1966}.

Such a form was found in cases where $H_1$ contains only one term of the form  $H_1 = x^m y^n$ \cite{Contopoulos1966a}.   
Further results were found in cases with two terms of degree three of the form \cite{Contopoulos_2025}
\begin{equation}  
H_1 = x y^2 + \alpha x^3, 
\end{equation}
where the perturbation is symmetric with respect to the axis $y=0$.
However, when we tried to apply this method to the most general perturbation of third degree in $x$ and $y$ 
\begin{equation}  
H_1 = \eta x y^2 + \alpha x^3+ \beta x^2 y + \gamma y^3,  \label{pert}
\end{equation}  
with  $\omega_1 = \omega_2=1$, we found three different secular terms containing $\sin(2t_0)t$, $\sin(t_0)t$ and $\sin(3t_0)t$ that could not be eliminated by using a second resonant integral beyond the zero-order Hamiltonian.  

{The problem that remained open until now was whether the same difficulty applies to other resonances in the general case of Eq.~\eqref{pert}. This is the main subject of the present work, where we conduct a detailed study of the problem involving several characteristic resonances. }

{It is worth noting that other, more recent techniques have also been developed to address the issue of secular terms arising at various resonances as: } 
\begin{enumerate}
    \item {the multiple time-scale analysis, which introduces multiple independent time variables to systematically capture both fast and slow dynamics, thereby eliminating secular terms and yielding uniformly valid approximations over extended time intervals \cite{kevorkian2012multiple,bender2013advanced}.
    \item the method of Lie normal forms where  the initial Hamiltonian is transformed to a simpler one via a series of canonical transformations \cite{giorgilli2022notes}.}
\end{enumerate}

{The Hamiltonian with the perturbation of Eq.~\eqref{pert} is appropriate in order  to evaluate the applicability of our approach for constructing formal integrals of motion and to explore various resonance phenomena. It has a rich nonlinear structure that enables the investigation of the onset of chaos and the identification of different resonance scenarios. At the same time, it represents a generalized form commonly encountered in molecular dynamics, where the harmonic oscillator approximation of intermolecular interactions is insufficient to describe anharmonic effects and nonlinear mode couplings \cite{Nordholm1974,Waite1981,Takatsuka2022}. Moreover it describes cubic couplings in galactic models that include as a special case the famous H\'{e}non-Heiles model for a star around a galactic center} \cite{Henon1964}).

The structure of the paper is the following: in Section 2 we give the general form of the integral when  in the perturbation of the form \eqref{pert} $\omega_1/\omega_2$  is irrational. In Section 3 we give the general form of the integral when $\omega_1/\omega_2=\frac{n}{m}$ and we apply it to two cases with $m=1$ ($n/m=4/1$ and $n/m=5/1$),  one case with $m=2$ ($n/m=3/2$) and a case with $m=3$ ($n/m=4/3$). In Section 4 we give the integral in the special cases $\omega_1/\omega_2=3/1, 2/1$ and $1/1$. We give also the forms of the invariant curves and of the orbits for various values of $\epsilon$ and of $\alpha,\beta,\gamma$ in the cases $\omega_1/\omega_2=2/1$ and $1/1$.  Finally, we give  an approximate irrational case with $\omega_1/\omega_2=5\sqrt{2}/7=1.010$ {which is close to the case $\omega_1/\omega_2=1$}. In Section 5 we draw our conclusions and in the Appendix A we provide some details about the computer programs used.

\section{Irrational ratio of frequencies}
The construction of a formal integral of motion  $\Phi$ starts from the requirement 
\begin{equation}\label{req}
\frac{d\Phi}{dt}=[\Phi, H] = 0,
\end{equation}
where 
\begin{equation}
    [\Phi,H]=\frac{\partial\Phi}{\partial x}\frac{\partial H}{\partial X}+\frac{\partial\Phi}{\partial y}\frac{\partial H}{\partial Y}-\frac{\partial\Phi}{\partial X}\frac{\partial H}{\partial x}-\frac{\partial\Phi}{\partial Y}\frac{\partial H}{\partial y}
\end{equation}
is the corresponding Poisson bracket.
{From Eqs.~\eqref{genham},\eqref{phit} we find  }\begin{equation}
[\Phi, H] = \left[ \sum_{q=0}^\infty \epsilon^q \Phi_q, \sum_{r=0}^\infty \epsilon^r H_r \right]
= \sum_{q=0}^\infty \sum_{r=0}^\infty \epsilon^{q+r} [\Phi_q, H_r]
= \sum_{s=0}^\infty \epsilon^s \sum_{k=0}^s [\Phi_k, H_{s-k}],
\end{equation}
{where we defined the total order $s = q + r$. Thus, for the  Poisson bracket to vanish at order $s=n+1$, each coefficient of $\epsilon^{n+1}$ must vanish independently:}
\begin{equation}
\sum_{k=0}^{n+1} [\Phi_k, H_{n+1-k}] = 0.
\end{equation}
This yields the recursive equation at order $n$:
\begin{equation}
[\Phi_n, H_0] + [\Phi_{n-1}, H_1] + \cdots + [\Phi_0, H_{n+1}] = 0,
\label{eqz}
\end{equation}
which determines $\Phi_n$ by requiring that $[\Phi, H] = 0$ holds up to order $\epsilon^{n+1}$, assuming that all lower order terms $\Phi_0, \ldots, \Phi_{n-1}$ as well $H_0, \ldots, H_{n+1}$ are known. Thus in the case of irrational frequencies  we start by inserting the  zero-order term $\Phi_{0} = \Phi_{1,0}=\frac{1}{2} \left( X^2 + \omega_1^2 x^2 \right)$ in Eq.~\eqref{eqz} and solve the system recursively.

 If we only have third order terms in the perturbation $H_1$ then there is an integral
\begin{equation}\label{int}    \Phi=\Phi_{0}+\epsilon\Phi_{1}+\epsilon^2\Phi_{2}+\dots
\end{equation}
and  Eq.~\eqref{eqz} becomes
\begin{equation}\label{diffeq}
    [\Phi_{n},H_0]=-[\Phi_{n-1},H_1].
\end{equation}
The solution of Eq.~\eqref{eqz} is found by using the zero-order solutions
\begin{eqnarray}
    \nonumber x=\frac{\sqrt{2\Phi_{1,0}}}{\omega_1}\sin(\omega_1(t-t_0)),\quad
    \nonumber y=\frac{\sqrt{2\Phi_{2,0}}}{\omega_2}\sin(\omega_2 t),\\
     X=\sqrt{2\Phi_{1,0}}\cos(\omega_1(t-t_0)),\quad
Y=\sqrt{2\Phi_{2,0}}\cos(\omega_2 t). \label{eqs}
\end{eqnarray}
Thus we express $[\Phi_{n-1},H_1]$ in terms of sines and cosines of $\omega_1(t-t_0)$ and $\omega_2t$ and then integrate Eq.~\eqref{diffeq} and find 
\begin{eqnarray}
    \Phi_n=-\int (\Phi_{n-1},H_1)dt,
\end{eqnarray}
where the integral contains trigonometric terms, that are then transformed to $x,y,X,Y$ by using again Eqs.~\eqref{eqs}.

A program that gives the form of the integral \eqref{int}
for non-commensurable frequencies $\omega_1,\omega_2$ is available up to high orders, but the result is too long to be printed here (8800 terms up to order 4 in $\epsilon$).

As an example we give here the form of $\Phi_1$
\begin{align}
\Phi_1=& \frac{4 \epsilon} {M}\bigg( x \big( y^2 \eta + x^2 \alpha + \frac{1}{2} x y \beta \big) \omega_1^4  \notag \\
&+ \big( \big(-\frac{9}{4} \omega_2^2 y^2 \eta  - 2 Y^2 \eta + \frac{1}{2} X Y \beta\big) x + \frac{1}{2} X y (4 Y \eta-X \beta )  -\frac{17}{4} \omega_2^2 x^3 \alpha - 2 \omega_2^2 x^2 y \beta 
\big) \omega_1^2  \notag \\
&+  \big( \big(\frac{1}{2} \omega_2^2 y^2 \eta + \frac{1}{2} Y^2 \eta- 2 X Y \beta \big) x  
+ 2 X y (- \frac{1}{4} Y \eta+X \beta )+ \omega_2^2 x^3 \alpha  \big) \omega_2^2\bigg),
\end{align}
where 
\begin{align}
    M=(\omega_1^2-4\omega_2^2)(4\omega_1^2-\omega_2^2).
\end{align}
In $\Phi_2$  we have the denominator 
\begin{align}
    \omega_1^4\omega_2^4(\omega_1^2-\omega_2^2)(\omega_1^2-4\omega_2^2)(4\omega_1^2-\omega_2^2)(\omega_1^2-9\omega_2^2)(9\omega_1^2-\omega_2^2)
\end{align}
and so on.

We see that in  these expressions the denominators contain terms of the form $(m^2\omega_1^2-n^2\omega_2^2)$. Resonances take place when $(m^2\omega_1^2-n^2\omega_2^2)=0$. 

The terms  $\Phi_N$ are of degree $N+2$ in $x,y,X,Y$. The denominators correspond in general to secular terms in the resonant cases $\omega_1=n,\omega_2=m$. However, there is an exception, namely the case $\omega_1=3,\omega_2=1$ (Section 4.1) , where the numerator of $\Phi_2$ has a factor $(9\omega_1^2-\omega_2^2)(\omega_1^2-9\omega_2^2)$ that eliminates the corresponding factor in the denominator. We note that the factor $(m^2\omega_1^2-n^2\omega_2^2)$ is always followed by the symmetric factor $(n^2\omega_1^2-m^2\omega_2^2)$. The factor $(\omega_1^2-\omega_2^2)$ does not appear in $\Phi_{1}$ but in $\Phi_{2}$ (see also Section 4.3).

\section{Integrals in general rational cases}
If  $\omega_1/\omega_2$ is rational
\begin{eqnarray}
    \frac{\omega_1}{\omega_2}=\frac{n}{m}
\end{eqnarray}
{with $\omega_1=n, \omega_2=m$ then  the construction of a formal integral is made according to the following steps:}
\begin{enumerate}
    \item {We set the zero-order integral }
    \begin{eqnarray}
    \substack{S_0 \\ C_0}=(2\Phi_{1,0})^{\frac{m}{2}}(2\Phi_{2,0})^{\frac{n}{2}}\substack{\sin\\ \cos}(mnt_0)t\label{zer}
\end{eqnarray}
{which is of degree $m+n$ in $x,y,X,Y$, where}
\begin{eqnarray}
    S_0^2+C_0^2=(2\Phi_{1,0})^m(2\Phi_{2,0})^{n}.
\end{eqnarray}
\item From $\substack{S_0\\C_0}$ we find formal integrals of the form
\begin{eqnarray}
    \nonumber S=S_0+\epsilon S_1+\epsilon^2S_2+\dots\\
    C=C_0+\epsilon C_1+\epsilon^2C_2+\dots
\end{eqnarray}
\item  In the terms $\epsilon^2S_2$ or $\epsilon^2C_2$ we find two secular terms $\substack{S_{2sec} \\ C_{2sec}}$ proportional to 
\begin{align}
    A=(2\Phi_{1,0})^{\frac{m+2}{2}}(2\Phi_{1,0})^{\frac{n}{2}}\substack{\sin \\ \cos}(mnt_0)t,\quad
    B=(2\Phi_{1,0})^{m/2}(2\Phi_{1,0})^{\frac{n+2}{2}}\substack{\sin \\ \cos}(mnt_0)t
\end{align}
of degree $m+n+2$ in $x,y,X,Y$.
\item We set a new zero-order integral of the form
\begin{align}
    \phi_0=c_1(2\Phi_{1,0})^2+2c_2(2\Phi_{1,0})(2\Phi_{2,0})+c_3(2\Phi_{2,0})^2\label{c1c2c3}
\end{align}
{ and calculate the formal integral
\begin{equation}
\phi_0+\epsilon\phi_1+\epsilon^2\phi_2+\dots
\end{equation}
up to degree $m+n+2$ in $x,y,X,Y$.} 

{We note here that the zero-order integral $\Phi_{1,0}$ gives the terms
\begin{equation}
\Phi_1=\Phi_{1,0}+\epsilon\Phi_{1,1}+\dots+\epsilon^{m+n-2}\Phi_{1,m+n-2},
\end{equation}
where $\Phi_{1,m+n-2}$ contains the secular terms $\Phi_{1sec}$ proportional to $\substack{\cos\\\sin}(mnt_0)t$\footnote{ this formula does not apply in the case $m=n=1$, see Section 4.3.}. In fact, $\Phi_{i,m+n-2}$ is of degree $m+n$ in $x,y,X,Y$. Then a zero-order integral $\Phi_{1,0}^2$ gives secular terms of the form $\epsilon^{m+n-2}(2\Phi_{1,0})\Phi_{1,m+n-2}$ which is of degree $m+n+2$ in $x,y,X,Y$.  } 

{If we would start an integral with $\Phi_{2,0}$ (instead of $\Phi_{1,0}$) then we would find the same secular terms with a minus sign ($-\Phi_{1sec}$) because the sum $\Phi_{1,0}+\Phi_{2,0}=H_0$ does not give any secular terms. Thus the more general case of a zero-order integral $\phi_0$ (Eq.~\eqref{c1c2c3}) has secular terms proportional to $\substack{\sin \\ \cos}(mnt_0)t,$ of degree $m+n-2$ in $x,y,X,Y$, i.e. $\phi_0$ gives also terms proportional to $A,B$.}

\item {Since $S_0$ or $C_0$ gives secular terms in $\substack{S_2 \\ C_2}$ of the form $\substack{\sin \\ \cos}(mnt_0)t$ of degree $m+n+2$ in $x,y,X,Y$ we  must multiply the zero-order integral $\substack{S_0 \\ C_0}$ by $\epsilon^{m+n-4}$ in order to match the terms of  degree $m+n+2$ in $x,y,X,Y$. Then  the secular terms generated by $\phi_{0}$ and by $\epsilon^{m+n-4}\substack{S_0 \\ C_0}$  are of the same degree and of the same form. 
Thus by  adding the corresponding series with an appropriate choice of $c_1,c_2,c_3$ we can eliminate all the secular terms of degree $m+n+2$.}
\end{enumerate}

If instead of $\omega_1=n$ and $\omega_2=m$ (where $m,n$ are the smallest integers forming the ratio $\omega_1/\omega_2$) we have $\omega_1 = k n$ and $\omega_2 = k m$, then by changing  the variables and the times ($t$ and $t_0$) by a factor $k$ we find the same equations of motion.

{It is important to note that the choice between  $S_0$ or $C_0$ as the zero-order integral depends on the parity of  $m+n$}. If $m+n$ is odd/even  the secular terms generated from $\phi_0$ contain the factor $\cos(mt_0)t/\sin(mt_0)t$ and the secular terms generated from $S_0$ when $m+n$ is odd/even contain the factor $\cos(mt_0)t/\sin(mt_0)t$. On the other hand, the secular terms generated from $C_0$ contain the factor $\sin(mt_0)t/\cos(mt_0)t$ when $m+n$ is odd/even. Therefore, in order to match the forms of the  secular terms and eliminate them by appropriate choices of $c_1,c_2,c_3$, we must use $S_0/C_0$ in the case $m+n=$odd/even.

The resulting integral has the form
\begin{align}
    \phi=\phi_0+\sum_{i=0}\epsilon^{m+n-4+i}\left(\phi_{m+n-4+i}+\substack{S_i\\C_i}\right),
\end{align}
where the secular terms of degree $m+n+2$ in $x,y,X,Y$ are eliminated. This formula is valid for $m+n>4$.  The special cases $m+n\leq4$ are described in Section 4.

In a similar way we eliminate the higher order secular terms, as in \cite{ContopMouts1966}.

In the following subsections (3.1-3.4) we give the explicit formulae for the secular terms in the following representative cases: two cases with $n=1$: $4/1$ ($m\cdot n = 4$, $m+n = 5=$ odd) and $5/1$ ($m\cdot n = 5$, $m+n = 6=$ even), one case with $m=2$: $3/2$ ($m\cdot n = 6$, $m+n = 5=$ odd) and one case with denominator $m=3$: $4/3$ ($m\cdot n = 12$, $m+n = 7=$ odd). Details about the calculations are given in the Appendix A.

We note that in our examples we work with $n>m$. Our results also apply in the symmetric cases with $\omega_1/\omega_2=n/m<1$ after replacing $x,y,X,Y$ by $y,x,Y,X$.

\subsection{Resonance $\omega_1/\omega_2=4/1$ ($\omega_1 = 4$, $\omega_2=1$)}

In this case we use the zero order integral
\begin{equation}
    S_0 = \left(2\Phi_{1,0}\right)^{1/2}\left(2\Phi_{2,0}\right)^{2}\sin\left(4t_0\right) = -4\,XY{y}^{3}-4\,x \left( {Y}^{4}-6\,{Y}^{2}{y}^{2}+{y}^{4} \right) +4
\,X{Y}^{3}y.
\end{equation}
{and we find that $S_1$ does not contain any secular terms. But $S_2$  contains the secular terms $S_{2sec}$  of order $m+n+2=7$ in $x,y,X,Y$} which are
\begin{align}
S_{2\text{sec}} = -\frac{15}{2048}\Bigg[&A\left(\frac{976}{945}\beta^2 - \frac{512}{5}\beta\gamma + \left(\alpha - \frac{16}{3}\eta\right)\left(\alpha - \frac{16}{15}\eta\right)\right) \notag \\
&+ \frac{32}{5}B\left(\alpha\eta - \frac{88}{9}\eta^2 + \frac{32}{189}\beta^2 + 16\beta\gamma - 640\gamma^2\right)\Bigg],
\end{align}
where 
\begin{equation}
    A=(2\Phi_{1,0})^{3/2}(2\Phi_{2,0})\cos(4t_0)t,\quad B=(2\Phi_{1,0})^{1/2}(2\Phi_{2,0})^2\cos(4t_0)t
\end{equation}.

Similar terms we find if we start with  $\phi_0=c_1(2\Phi_{1,0})^2+2c_2(2\Phi_{1,0})(2\Phi_{2,0})+c_3(2\Phi_{2,0})^2$:
\begin{align}
\phi_{3\text{sec}} =\frac{1}{384}
&\left(A\left(-2\eta^2 + \alpha\eta + 120\left(\gamma - \frac{1}{5}\beta\right)\left(\gamma - \frac{2}{21}\beta\right)\right)\right)\eta(2c_1 - c_2) \notag \\
&+ \left(B\left(-2\eta^2 + \alpha\eta + 120\left(\gamma - \frac{1}{5}\beta\right)\left(\gamma - \frac{2}{21}\beta\right)\right)\right)\eta(c_2 - 2c_3).\label{f3sec}
\end{align}
In this case $m+n-4=1$, therefore the secular terms are of degree $m+n+2=7$ in $x$, $y$, $X$, $Y$, both in $\phi_3$ and $S_2$. The formal integral is
\begin{align}
\phi=\phi_0+\epsilon(\phi_1+S_0)+\epsilon^2(\phi_2+S_1)+\epsilon^3(\phi_3+S_2)
\end{align}
and therefore $S_0$ has to be multiplied by $\epsilon^{m+n-4}=\epsilon$ in order to match the power of $\epsilon$ in $\phi_1$. Then by setting
\begin{align}
   \nonumber&c_1=\frac{1}{1344}  \frac{945\alpha^2 - 6048\alpha\eta + 976\beta^2 - 96768\beta\gamma + 5376\eta^2}{\left(-2\eta^2 + \alpha \eta + \frac{16}{7}(\beta - 5\gamma)\left(\beta - \frac{21}{2}\gamma\right)\right) \eta},\\&\nonumber c_2=0,\\&
   c_3=\frac{1}{42}  \frac{-189\alpha \eta - 32\beta^2 - 3024\beta\gamma + 120960\gamma^2 + 1848\eta^2}{\left(-2\eta^2 + \alpha \eta + \frac{16}{7}(\beta - 5\gamma)\left(\beta - \frac{21}{2}\gamma\right)\right) \eta},\label{cs}
\end{align}
we eliminate the secular terms of degree 7 in $x,y,X,Y$. Similar calculations are expected for terms of higher order as it was done in our paper \cite{ContopMouts1966}.

On the other hand, if we consider $C_0$ as a zero-order term, we find in $C_2$ terms with $\sin(2t_0)t$ and these cannot be eliminated by using $\phi_0$ as zero order term.

In the particular case $\eta=0$ the formulae \eqref{cs} cannot be applied, because then $c_1$ and $c_3$ go to infinity. This is due to the fact that the secular terms of degree 7, as derived from the zero-order integral $\phi_0$, are zero (Eq.~\eqref{f3sec}). But the secular terms of degree 9 are different from zero. Then in this particular case we must combine $S_0$ with $\phi_3$ in order to eliminate the secular terms in $\phi_5$.

In the more peculiar cases $\eta=\beta=0$ the Hamiltonian is separable and thus there are no secular terms at all. A summary of the above cases is given in Table~\ref{t1}.

\begin{table}[H]
\centering
\begin{tabular}{|p{4cm}|p{4cm}|p{5cm}|}
\hline
{Case} & {Secular Terms } & {Can They Be Eliminated?} \\
\hline
$S_0$, arbitrary $\eta, \alpha, \beta, \gamma$
\newline with $\eta\beta\neq$ 0& 
Secular terms \newline 
of degree 7 appear \newline 
in $S_2$ & 
Yes, by using appropriate 
$c_1, c_2, c_3$ in $\phi_0$.\\
\hline
$S_0$, $\eta=0$, $\beta \neq 0$. & 
Secular terms 
of degree 9 appear
in $S_2$. & 
The secular terms of degree 7 vanish while those of degree-9  can 
be eliminated by using 
$S_0$ and $\phi_3$ and appropriate $c_1, c_2, c_3$. \\
\hline
$S_0$, $\eta=0$, $\beta=0$ & 
No secular terms. & 
The Hamiltonian 
is separable. \\
\hline
$C_0$, arbitrary parameters & 
Secular terms appear 
in $C_2$. & 
No they 
cannot be eliminated. \\
\hline
\end{tabular}
\caption{Presence and eliminability of secular terms in the case $\omega_1/\omega_2 = 4/1$.}\label{t1}
\end{table}

\subsection{Resonance $\omega_1/\omega_2=5/1$ ($\omega_1 = 5$, $\omega_2=1$)} 

In this case the zero-order integral is 
\begin{align}
\nonumber C_0&=(2\Phi_{1,0})^{1/2}(2\Phi_{2,0})^{5/2}\cos(5t_0)\\&=X{Y}^{5}-10\,X{Y}^{3}{y}^{2}+5\,XY{y}^{4}+25\,{Y}^{4}xy-50\,{Y}^{2}x{y
}^{3}+5\,x{y}^{5}.
\end{align}
We have secular terms proportional to  $A=(2\Phi_{1,0})^{3/2} (2\Phi_{2,0})^{3/2}\sin(5t_0)t$
and \\$B=(2\Phi_{1,0})^{1/2}(2\Phi_{2,0})^{7/2}\sin(5t_0)t$ in the terms of order 6 in $x,y,X,Y$ both from  $\phi_{0}$ and  $C_0$. The secular terms are not printed here.

In this case $m+n-4=2$. The secular terms of the integral starting with zero-order $C_0$ are  of degree $m+n-2=8$ and the secular terms of the integral starting with $\phi_0$ are also of degree 8. Therefore $C_0$ has to be multiplied by $\epsilon^{m+n-4}=\epsilon^2$ in order to eliminate the secular terms by taking appropriate values of $c_1,c_2,c_3$ in a formal integral 
\begin{equation}
\phi = \phi_0 + \epsilon \phi_1 + \epsilon^2 \left(\phi_2 + C_0\right) + \epsilon^3 \left(\phi_3+C_1\right) + \epsilon^4 \left(\phi_4 + C_2 \right).
\end{equation}
The forms of the secular terms and of $c_1,c_2,c_3$ are similar to those of the case $\omega_1/\omega_2=4/1$ with the change of $\epsilon S_0$ by $\epsilon^2C_0$ because now $m+n=6$ (even).

On the other hand, if we take $S_0$ instead of $C_0$, we find secular terms of the form
$\sin(5t_0)t$ and these cannot eliminate the secular terms due to $\phi_0$.


\subsection{Resonance $\omega_1/\omega_2=3/2$ ($\omega_1 = 3$, $\omega_2=2$)}

In this case we have
\begin{equation}
    S_0 =(2\Phi_{1,0})(2\Phi_{2,0})^{3/2}\sin(6t_0)=6\,{X}^{2}{Y}^{2}y-8\,{X}^{2}{y}^{3}-6\,X{Y}^{3}x+72\,XYx{y}^{2}-54\,{
Y}^{2}{x}^{2}y+72\,{x}^{2}{y}^{3}.
\end{equation}
and the secular terms in $S_2$ and $\phi_3$ are proportional to 

\begin{align}
    A = \left(2\Phi_{1,0}\right)^2 \left(2\Phi_{2,0}\right)^{3/2} \cos\left(6t_0\right)t,\quad 
    B = \left(2\Phi_{1,0}\right)\left(2\Phi_{2,0}\right)^{5/2} \cos\left(6t_0\right)t.
\end{align}
In this case $m+n-2=3$, therefore $S_0$ has to be multiplied by $\epsilon^{m+n-4}=\epsilon$ and
we can eliminate the secular term of degree 7 in $x,y,X,Y$ if we add $\phi_0 + \epsilon \phi_1 + ...$ and $\epsilon\left(S_0 + ...\right)$ and use appropriate values for $c_1,c_2,c_3$.
Namely,
\begin{align}
    \phi=\phi_0+\epsilon(\phi_1+S_0)+\epsilon^2(\phi_1+S_1)+\epsilon^3(\phi_2+S_2).
\end{align}

The secular terms and the values of $c_1, c_2, c_3$ that eliminate them are similar to those of the case $\omega_1/\omega_2=4/1$. The difference of this case is in the form of $S_0$ and $A,B$.


\subsection{Resonance $\omega_1/\omega_2=4/3$ ($\omega_1 = 4$, $\omega_2=3$)}

In this case we find two secular terms of degree $m+n+2=9$ in $x, y, X, Y$, namely proportional to
\begin{align}
    A = \left(2\Phi_{1,0}\right)^{5/2} \left(2\Phi_{2,0}\right)^2 \cos\left(12t_0\right)t,\quad
    B = \left(2\Phi_{1,0}\right)^{3/2} \left(2\Phi_{2,0}\right)^3 \cos\left(12t_0\right)t.
\end{align}
These appear when the zero-order terms are either 
\begin{align}
    \nonumber&S_0 = (2\Phi_{1,0})^{3/2} (2\Phi_{2,0})^{2}\sin(3t_0)=
 16x^3\left(Y^4 - 54Y^2y^2 + 81y^4\right)\sqrt{16} \\&
 + 12X\Big[9Yy^3\left(16x^2 - \frac{1}{3}X^2\right)\sqrt{9} 
- Xx\left(Y^4 - 54Y^2y^2 + 81y^4\right) 
- 3\left(16x^2 - \frac{1}{3}X^2\right)yY^3\Big]
\end{align}
or $\phi_0$ (Eq.~(24)).
In this case we have $m+n-4=3$, therefore the term $S_0$ has to be multiplied by $\epsilon^3$. Then we have 
\begin{align}
\phi=\phi_0+\epsilon\phi_1+\epsilon^2\phi_2+\epsilon^3(\phi_3+S_0)+\epsilon^4(\phi_4+S_1)+\epsilon^5(\phi_5+S_2)
\end{align}
up to degree $m+n+2=9$ in $x,y,X,Y$.

The solutions for $c_1$, $c_2$, $c_3$ that eliminate the secular terms using $\eta$, $\alpha$, $\beta$, $\gamma$ are lengthy and they are not given here.


\section{Special Cases ($m+n\leq 4$)}

\subsection{Resonance $\omega_1/\omega_2=3/1$ $(\omega_1 = 3,\, \omega_2=1)$}

In this case the zero order terms are $\phi_0 + C_0$
where
\begin{align}
    C_0=(2\Phi_{1,0})^{1/2}(2\Phi_{2,0})^{3/2}\cos(3t_0)=X Y^3 - 3 X Y y^2 + 9 Y^2 x y - 3 x y^3.
\end{align}
{This term must be multiplied by  $\epsilon^{m+n-4}=\epsilon^0=1$}. Thus $C_0$ and $\phi_0$ are of the same degree (zero) in $\epsilon$. The first order term $\phi_1$, after $\phi_0$, has no secular terms. Secular terms appear in $\phi_2$  and they are all proportional to $\sin(3t_0)t$ namely
\begin{align}
    \nonumber\phi_{2sec}=
A \left( -\eta \left( \gamma - \frac{1}{5} \beta \right) \left( c_1 - \frac{1}{2} c_2 \right) \right)
+ B \left( -\eta \left( \gamma - \frac{1}{5} \beta \right) \left( c_2 - 2 c_3 \right) \right),
\end{align}
where 
\begin{align}
    &A=(2\Phi_{1,0})^{3/2}(2\Phi_{2,0})^{3/2}\sin(3t_0)t,\quad
    B=(2\Phi_{1,0})^{1/2}(2\Phi_{2,0})^{5/2}\sin(3t_0)t.
\end{align}
{Similar secular terms we find if we start with the zero-order integral $C_0$. Then the formal integral reads:}
\begin{align}
    \phi=\phi_0+C_0+\epsilon(\phi_1+C_1)+\epsilon^2(\phi_2+C_2).
\end{align}
Therefore if we take
\begin{align}
    &\nonumber c_1=\frac{1}{2268} \frac{-175 \alpha^2 + 630 \alpha \eta - 99 \beta^2 + 5670 \beta \gamma - 756 \eta^2}{\eta (-5 \gamma + \beta)},\\&\nonumber 
    c_2=0,\\&
    c_3=\frac{1}{252} \frac{70 \alpha \eta + 12 \beta^2 + 630 \beta \gamma - 483 \eta^2 - 14175 \gamma^2}{\eta (-5 \gamma + \beta)}
\end{align}
we can  eliminate the secular terms. However if, instead of $C_0$, we start with zero-order integral $S_0=3XYy^2-Xy^3-3Y^3x+9Xxy^2$  the secular terms of $S_2$ are proportional to $\cos(3t_0)t$ {and they do not match with the those due to $\phi_{0}$}.

The present case is peculiar because the denominators of $c_1$ and $c_3$ contain only $\beta$ and $\gamma$ besides $\eta$ and tend to zero if $\beta$ and $\gamma$ tend to zero. If $\beta=\gamma=0$ ($\eta \alpha\neq 0$) the secular terms of $\phi_0$ are zero. This is due to the fact that the formal integral
\begin{eqnarray}    \Phi=\Phi_{1,0}+\epsilon\Phi_{1,1}+\epsilon^2\Phi_{12}+\dots
\end{eqnarray}
has no secular terms. In fact, we have checked that in the general form of the non-resonant integral  the denominator is eliminated when $\omega_1/\omega_2=3/1$ by a similar factor $(9\omega_1^2-\omega_2^2)(\omega_1^2-9\omega_2^2)$ in the numerator if $\beta=\gamma=0$.

Therefore, when  $\eta\alpha\neq 0$ and $\beta=\gamma=0$, there is no secular term in $\Phi$ and the zero-order formal integral of motion is not a combination of $C_0$ with $\phi_{0}$, but $\Phi_{1,0}$ itself. The same result is reached if $\eta=\alpha=0$ but $\beta\gamma\neq0$. We have checked that in the term $\Phi_{1,4}$ there are no secular terms, and probably there are no secular terms of even higher order. Nevertheless, the integral $\Phi$ is only formal and when the perturbation $\epsilon$ is strong we find chaos.


\subsection{Resonance $\omega_1/\omega_2=2/1$ $(\omega_1 = 2,\, \omega_2=1)$}

This case is quite special because we have to use a zero-order integral 
\begin{eqnarray}
S_0=(2\Phi_{1,0})^{1/2}(2\Phi_{2,0})\sin(2t_0)=2x(y^2-Y^2)+2yXY.
\end{eqnarray}
{We find no secular terms in $S_1$ but two secular terms} in $S_2$
\begin{align}
    \nonumber S_{2sec}=A \left( \frac{8}{5} \alpha \eta - \frac{32}{15} \eta^2 + \frac{64}{225} \beta^2 + \frac{32}{5} \gamma \beta - 64 \gamma^2 \right)\\
+ B \left( - \frac{32}{5} \gamma \beta - \frac{8}{5} \alpha\eta - \frac{4}{15} \eta^2 + \alpha^2 + \frac{52}{225} \beta^2 \right),
\end{align}
where 
\begin{eqnarray}
    A=(2\Phi_{1,0})^{\frac{3}{2}}(2\Phi_{2,0})\cos(2t_0)t,\quad
    B=(2\Phi_{1,0})^{\frac{1}{2}}(2\Phi_{2,0})^2\cos(2t_0)t.
\end{eqnarray}
On the other hand, if we start with the zero-order integral $\phi_0$ we find in $\phi_1$ the secular terms
\begin{eqnarray}
    \phi_{1sec}= -\eta \left[ A\left(c_1 - \frac{1}{2}c_2\right) + B\left( \frac{1}{2}c_2 - c_3 \right) \right].
\end{eqnarray}
Thus if we combine $S_0+\epsilon S_1+\dots$ and the terms $\epsilon \phi_0+\epsilon^2\phi_1+\dots$ .with 
\begin{align}
   \nonumber &c_1={\frac {-225\,\alpha^{2}+360\,\alpha\eta-52\,{\beta}^{2}+1440\,\gamma\,\beta+
60\,{\eta}^{2}}{1920\,\eta}},\\&\nonumber
   c_2=0,\\&
    c_3={\frac {45\,\alpha\eta+8\,{\beta}^{2}+180\,\gamma\,\beta-60\,{\eta}^{2}-
1800\,{\gamma}^{2}}{240\,\eta}}
\end{align}
the total secular terms are zero.
Then the formal integral is
\begin{eqnarray}
    \phi=S_0+\epsilon(S_1+\phi_0)+\epsilon^2(S_2+\phi_1)
\end{eqnarray}
and does not have secular terms of degree $5$ in $x,y,X,Y$.
The higher-order secular terms of this integral are expected to be eliminated in the same way as in \cite{ContopMouts1966}.
{We note that this formula is valid if $\eta\neq 0$ and $\beta$ and -or $\gamma$ are different from zero. If $\beta=\gamma=0$ it is valid even when $\eta=0$.}
\begin{figure}[H]
    \centering    \includegraphics[width=0.4\linewidth]{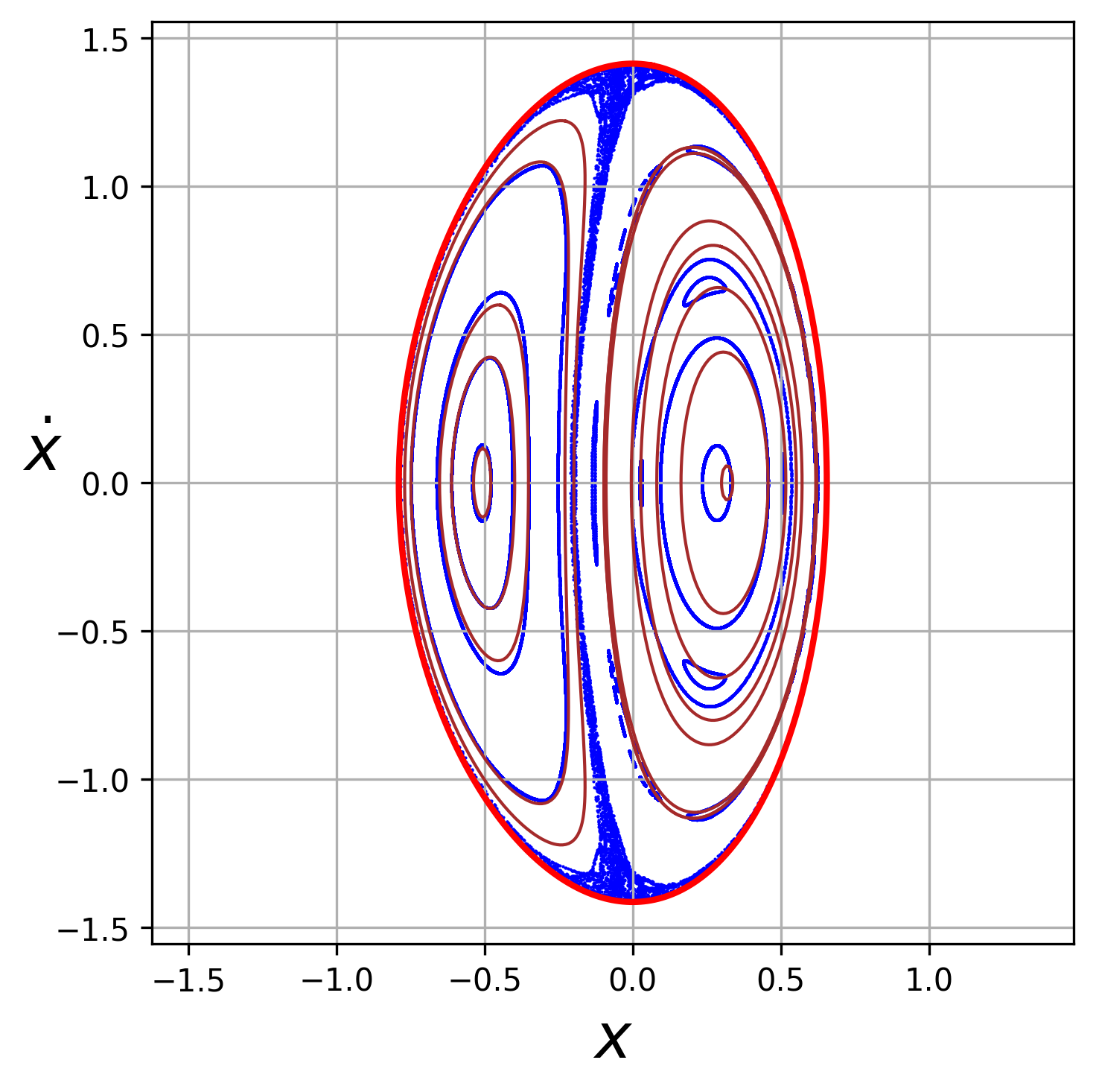}[a]
    \includegraphics[width=0.4\linewidth]{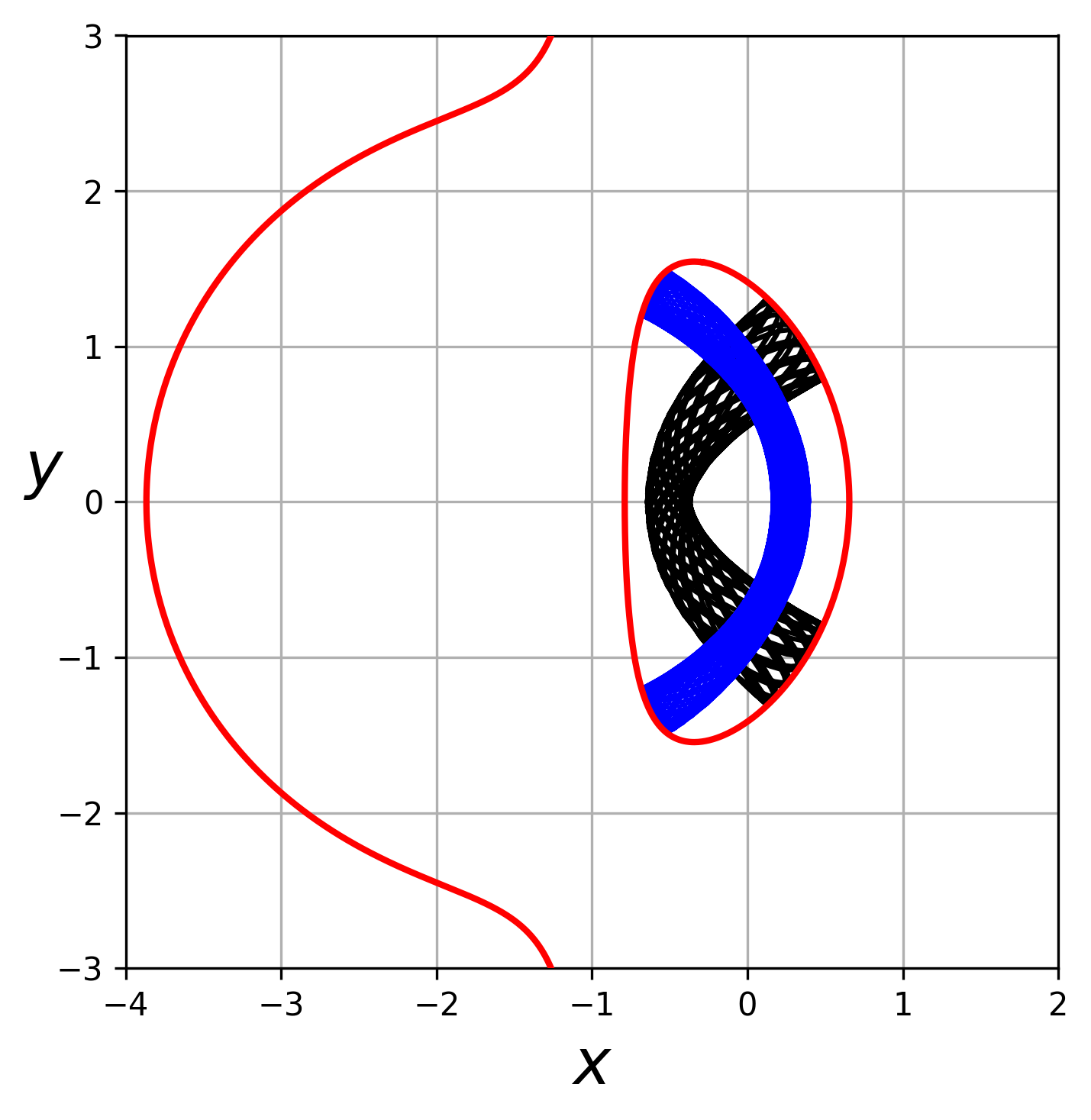}[b]\\
     \includegraphics[width=0.47\linewidth]{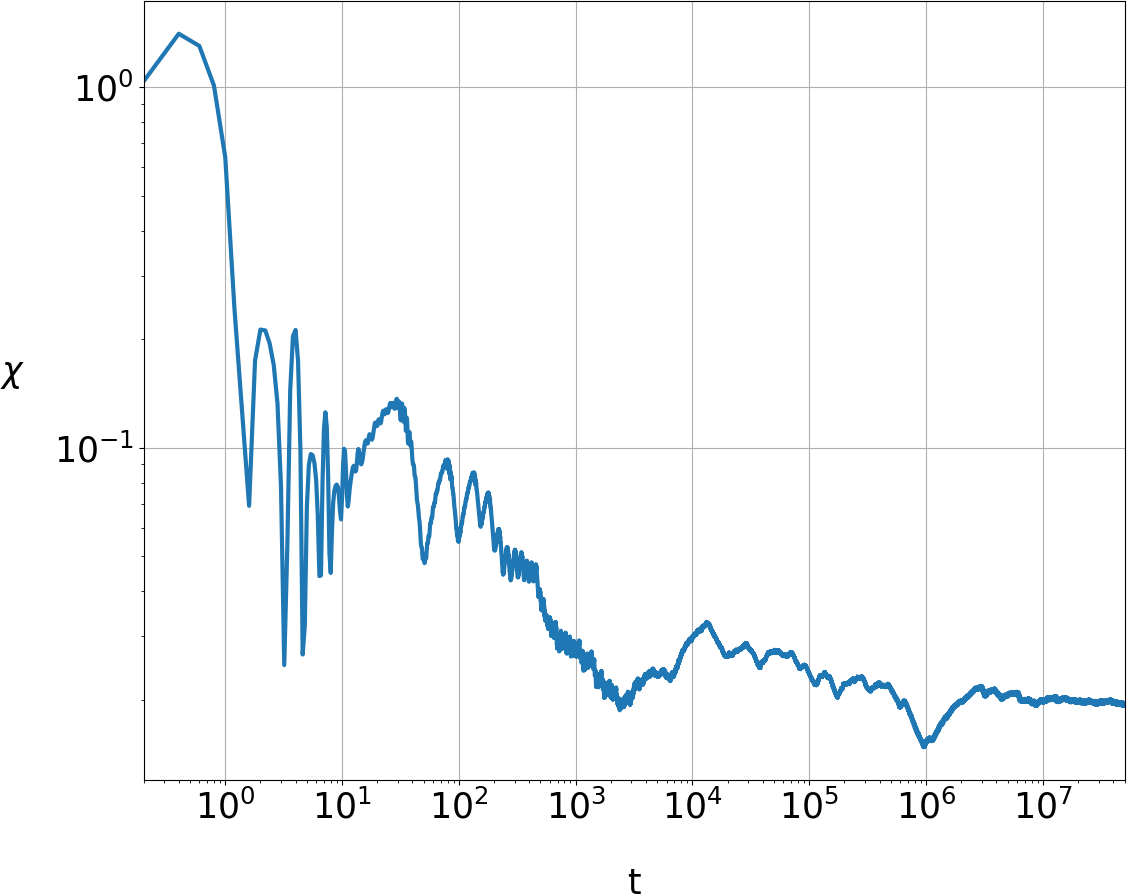}[c]
    \caption{Case $\epsilon = 0.5$, $\eta=\alpha=1$, $\beta=\gamma=0$, $\omega_1=2$, $\omega_2=1$. (a) Invariant curves, theoretical (red) and exact (blue). The blue region contains chaotic orbits. (b) Lissajous orbits corresponding to the left part of (a) (black) and to the right part (blue). The red curve is the curve of zero velocity (CZV). c) The finite time Lyapunov characteristic number for an orbit in the chaotic sea of a) ($x(0)=-0.0676, y(0)=0.0, p_x(0)=1.36, p_y(0)=0.364,\delta x(0)=1,\delta y(0)=0,\delta p_x=0,\delta p_y=0, E=1$). {We see that it saturates at the value $\chi\simeq 0.02$.}}
    \label{fig:r1}
\end{figure}

We calculated the invariant curves and the orbits for various values of $\epsilon$ and $\beta,\gamma$. If we set $\epsilon=0.5, \alpha=1,\beta=\gamma=0,\omega_1=2,\omega_2=1$ we find the surface of section $y=0$ of \figref{r1}a. The red curves are the invariant curves given by the integral $\phi$ (truncated at order $\epsilon^2$) and the blue curves are the corresponding exact invariant curves found by integrating the equations of motion. We see that the red and the blue curves are very close to each other. However, there are also some chaotic orbits (blue regions close to the axis $x=0$ with concentrations on the top and on the bottom). The ordered orbits in this case are generalized Lissajous figures (\figref{r1}b). 

{In \figref{r1}c we show the `finite time Lyapunov characteristic number' $\chi$ for an orbit inside the chaotic sea of  \figref{r1}a.
We remind that if we take two  nearby trajectories at $t=t_0$ and their deviation vector $\xi_k$ at the times $t=s\Delta t, s=1,2,\dots$ then $\chi$ is defined as }
\begin{equation}\label{chi}
\chi=\frac{1}{t}\sum_{i=1}^sa_i,
\end{equation}
where 
\begin{equation}
a_s=\ln\left|\frac{\xi_{s+1}}{\xi_s}\right|
\end{equation}
{is the `stretching number'. 
The $LCN$ itself is the limit of $\chi$ at $s\to\infty$ and is zero for ordered orbits, while it is a positive constant for chaotic orbits. }

\begin{figure}[H]
    \centering
    \includegraphics[width=0.4\linewidth]{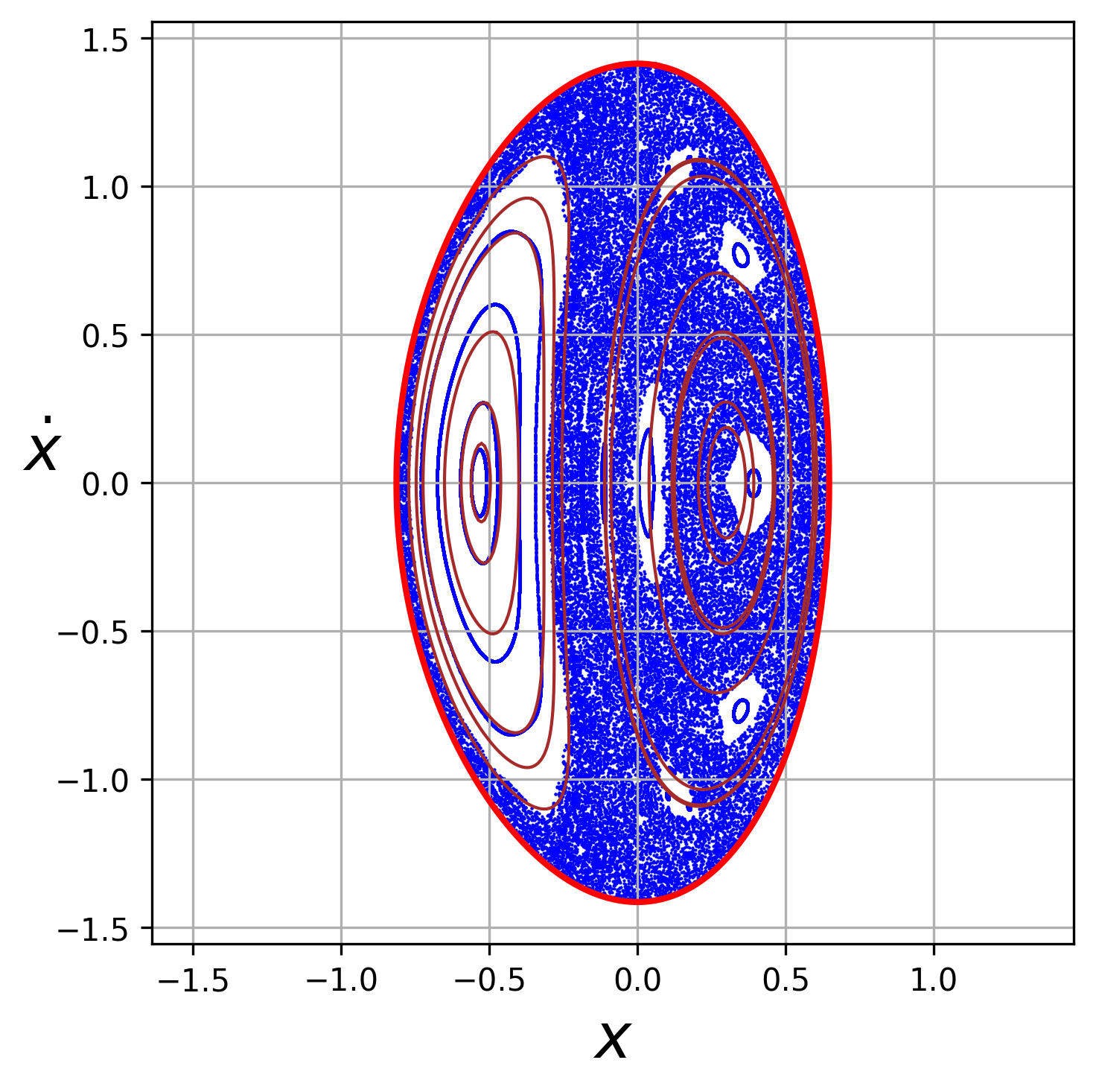}[a]
    \includegraphics[width=0.395\linewidth]{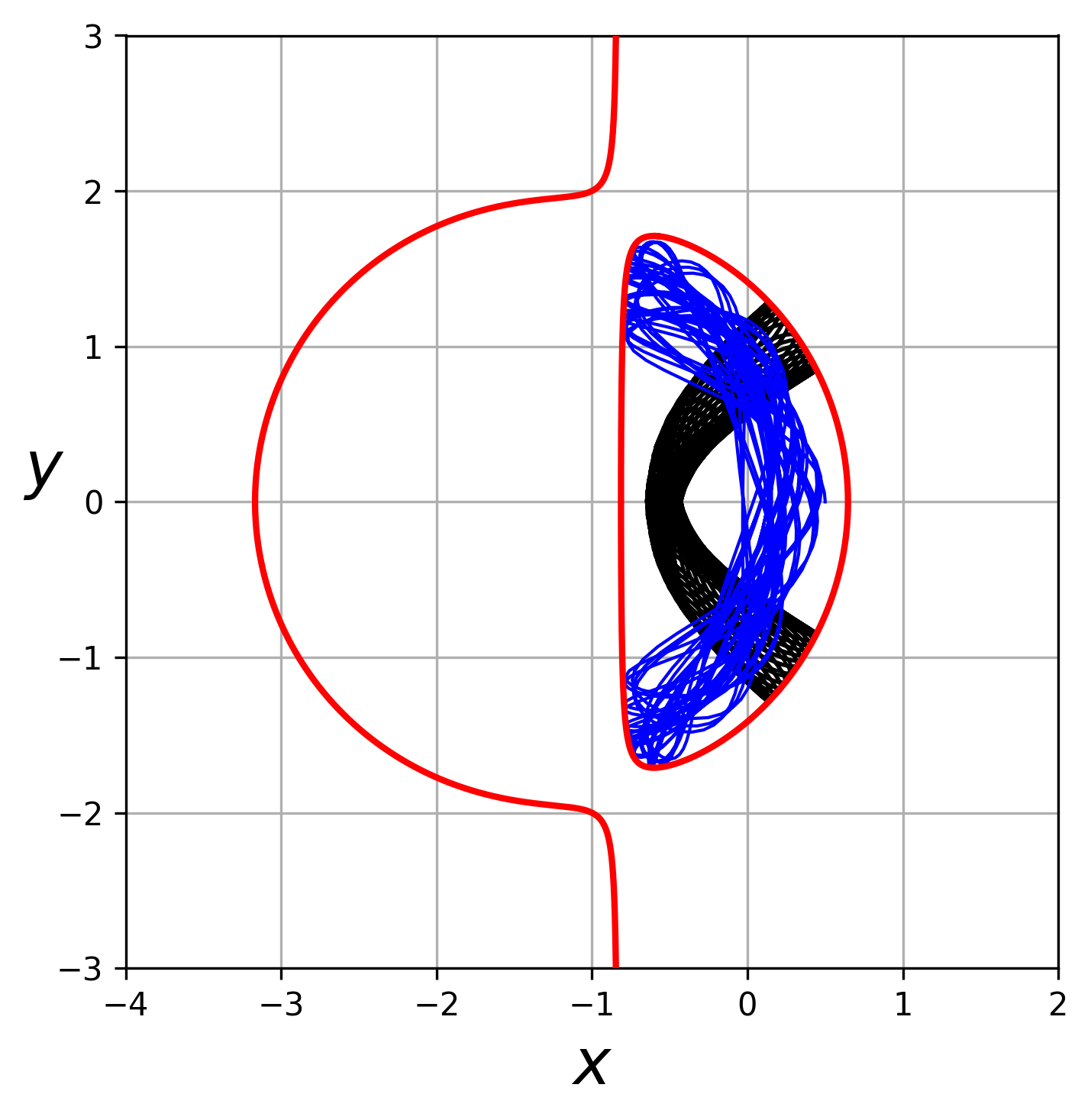}[b]
     \includegraphics[width=0.47\linewidth]{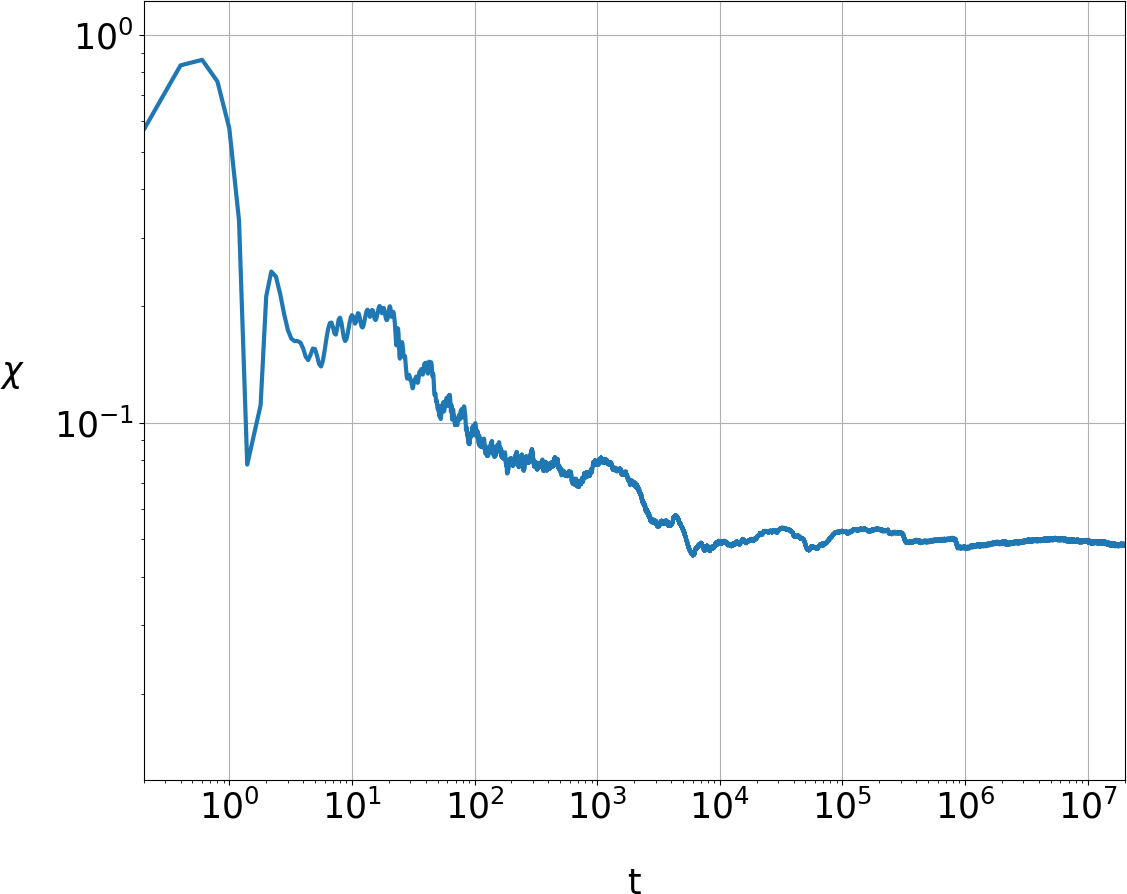}[c]
    \caption{Case $\epsilon = 0.6$, $\eta=\alpha=1$, $\beta=\gamma=0$, $\omega_1=2$, $\omega_2=1$. a) The corresponding Poincar\'{e} surface of section. The right part of a) contains mostly chaotic orbits. b)  Two representative orbits. c) The finite time Lyapunov characteristic number for an orbit in the chaotic sea of a) ($x(0)=-0.25, y(0)=0.0, p_x(0)=0, p_y(0)=1.323,\delta x(0)=1,\delta y(0)=0,\delta p_x=0,\delta p_y=0, E=1$). {We see that it saturates at the value $\chi\simeq 0.051$.}}
    \label{fig:r2}
\end{figure}

If we take a smaller $\epsilon$ the agreement is better and the chaotic domains are smaller while if we increase $\epsilon$ chaos becomes dominant. When $\epsilon=0.6$ (with the other constants the same), the right part of \figref{r2}a is mostly chaotic, with only some islands of stability. The red curves on the right (theoretical integral curves) do not represent any real invariant curves. However, on the left, the theoretical invariant curves (red) are very close to the exact curves (blue). The forms of the orbits in this case are shown in \figref{r2}b. The orbits starting on the left (for $y=0$) are generalized Lissajous figures (black), while those starting on the right (blue) are chaotic in general, but for a rather long time they look like Lissajous figures. {The corresponding $\chi$ for an orbit inside the chaotic sea of \figref{r2}a is shown in \figref{r2}c. We see that it is larger than that of \figref{r1}c.}

\begin{figure}[H]
    \centering
    \includegraphics[width=0.4\linewidth]{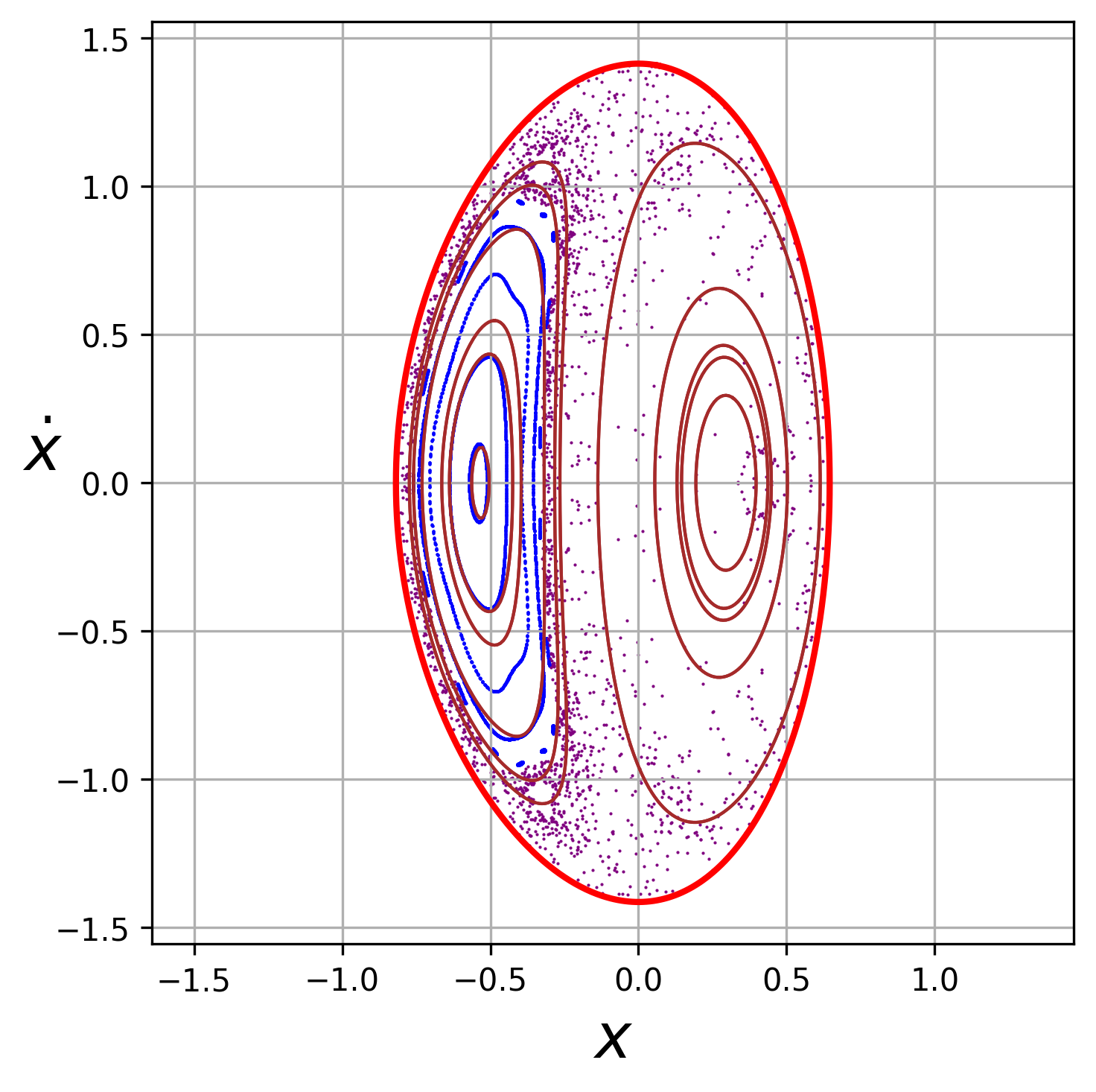}[a]
    \includegraphics[width=0.39\linewidth]{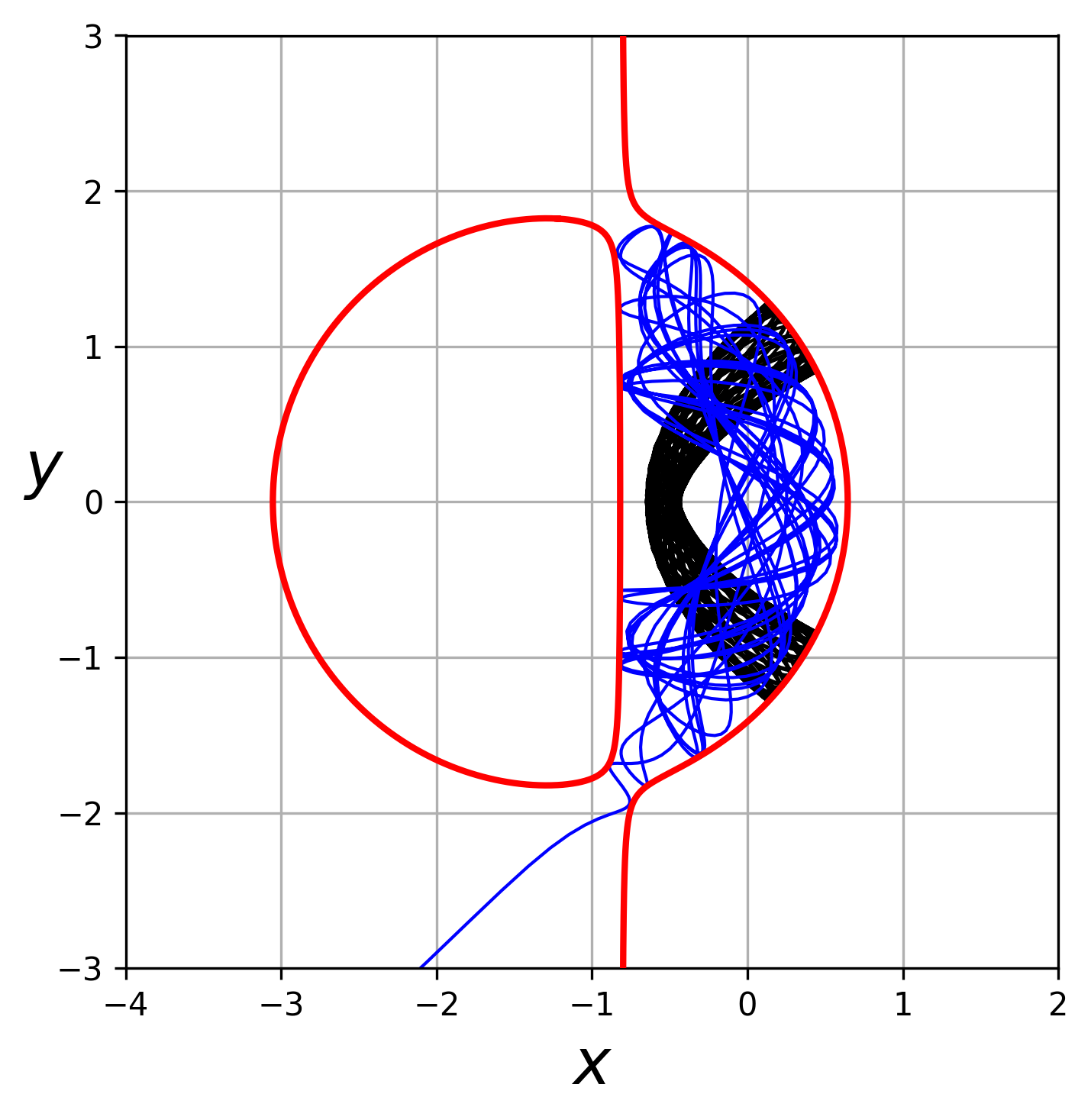}[b]
    \caption{Case $\epsilon = 0.62$, $\eta=\alpha=1$, $\beta=\gamma=0$, $\omega_1=2$, $\omega_2=1$.  a) The corresponding  Poincar\'{e} surface of section. b) Two characteristic orbits. Just above the energy of  escape along $y^2$  many chaotic orbits in b) (blue) escape to $y=\pm\infty$ through the openings of the CZV.}
    \label{fig:r3}
\end{figure}

Then for $\epsilon=0.62$ (the other constants are the same) we have escapes towards $y^2=\infty$. The invariant curves (\figref{r3}a) on the left are similar to the theoretical ones, but on the right most orbits escape (dots) and are not restricted  (\figref{r3}b) by the theoretical invariant curves (red). The corresponding orbits  starting on the left for $y=0$ (black) are of Lissajous form while those starting on the right (blue) are chaotic for some time and then they escape to $y^2=\infty$, passing from an opening of the CZV (upwards or downwards).

Then we consider the orbits of a fully perturbed case with $\eta=\alpha=\beta=\gamma=1$ ($\omega_1=2,\omega_2=1$).
Topologically the forms of the invariant curves and of the orbits are similar with those of the case $\eta=\alpha=1,\beta=\gamma=0)$ but numerically they are very different.

\begin{figure}[H]
    \centering
    \includegraphics[width=0.4\linewidth]{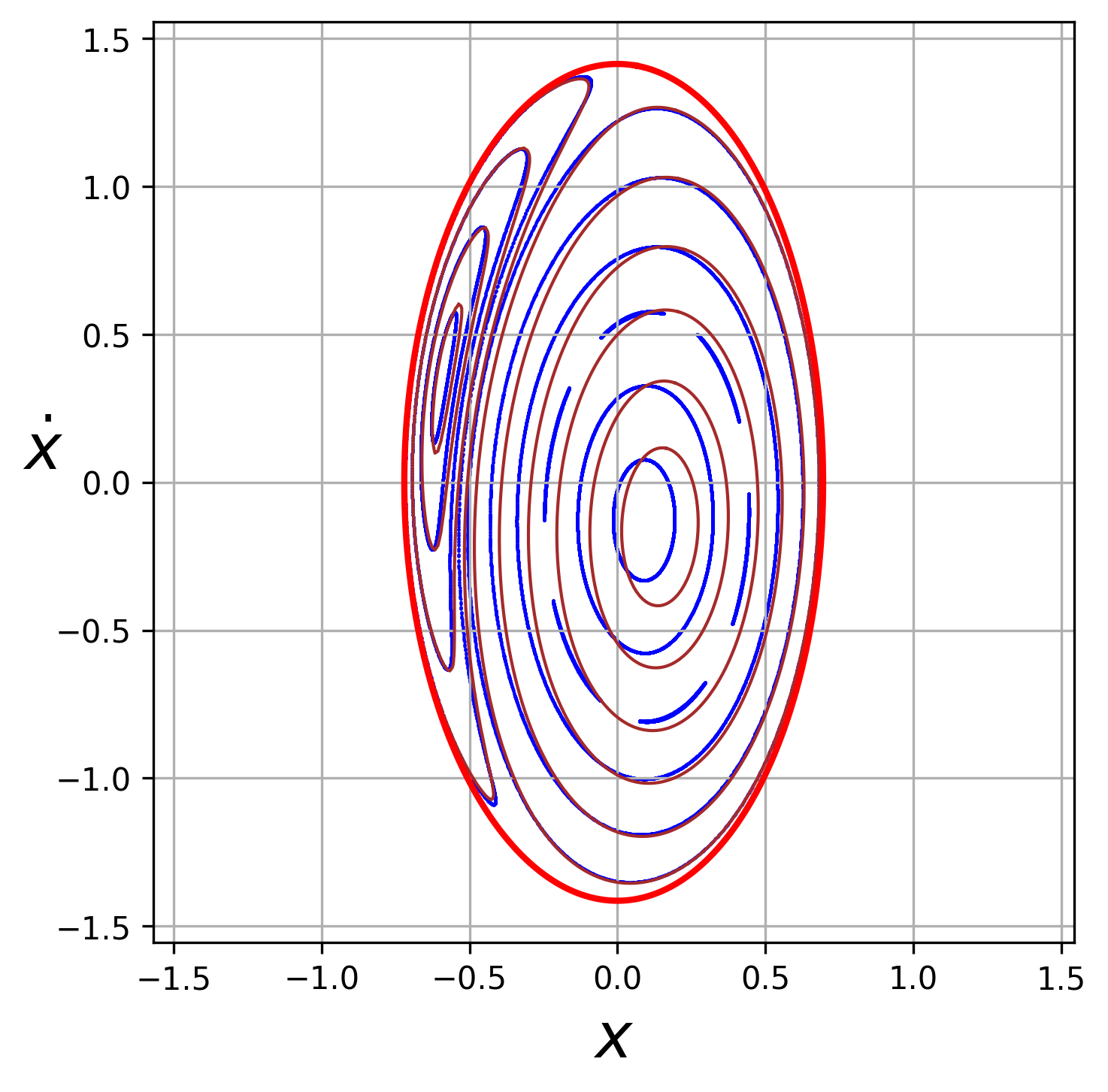}[a]
    \includegraphics[width=0.395\linewidth]{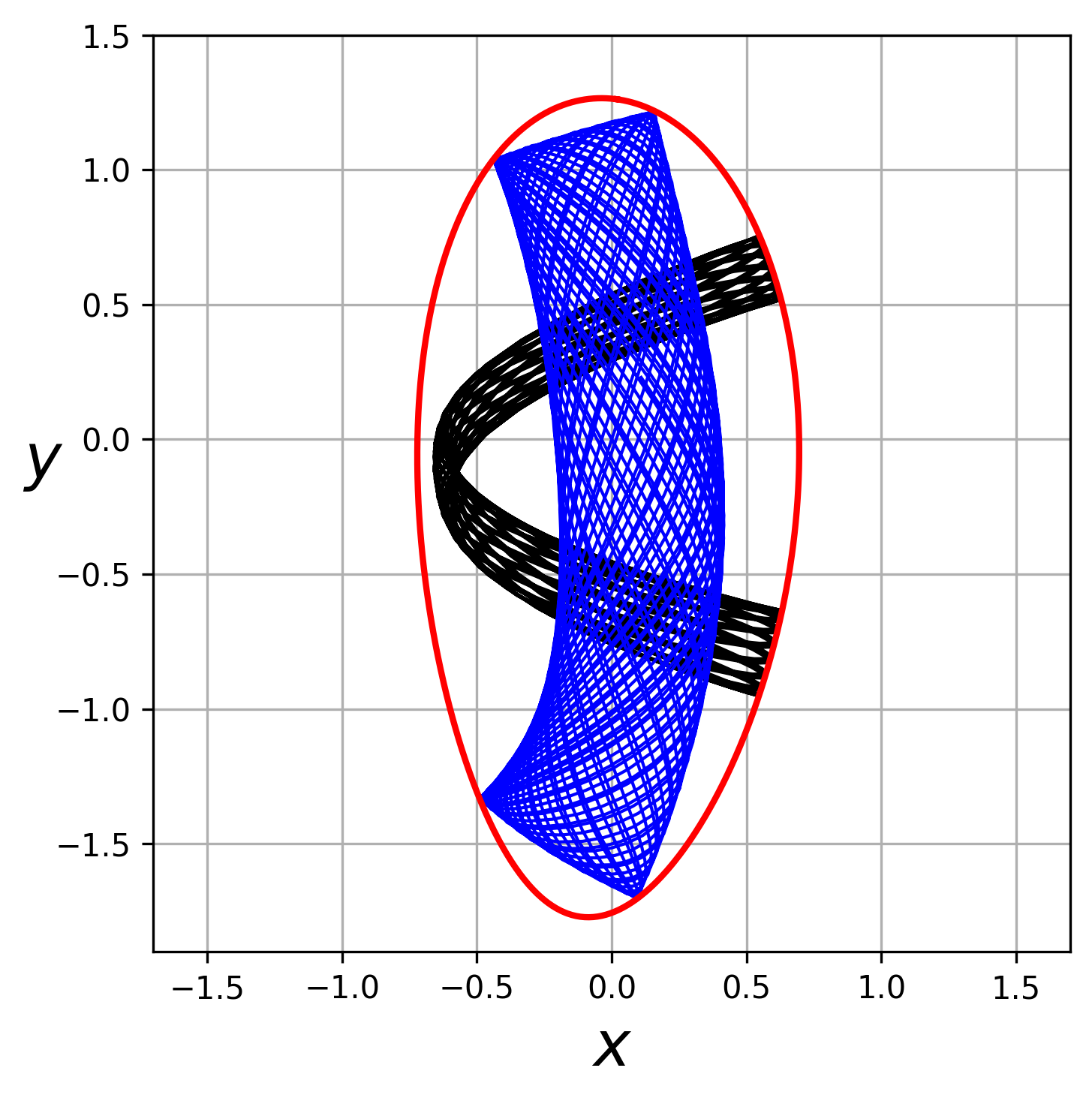}[b]
    \caption{Case with $\epsilon = 0.1$, $\eta=\alpha=\beta=\gamma=1$, $\omega_1=2$, $\omega_2=1$. (a)\ The theoretical invariant curves (red) and the corresponding exact (numerical) curves (blue). (b) Two representative Lissajous orbits.}
    \label{fig:r4}
\end{figure}

For $\epsilon=0.1$ we have invariant curves both on the right and on the left (\figref{r4}a) but the left curves form a small set and they are almost symmetric as in the case of \figref{r1}a. The theoretical invariant curves (red) are very similar with the exact invariant curves (blue). The corresponding orbits (\figref{r4}b) are Lissajous curves as in the case of \figref{r1}b.

\begin{figure}[H]
    \centering
    \includegraphics[width=0.4\linewidth]{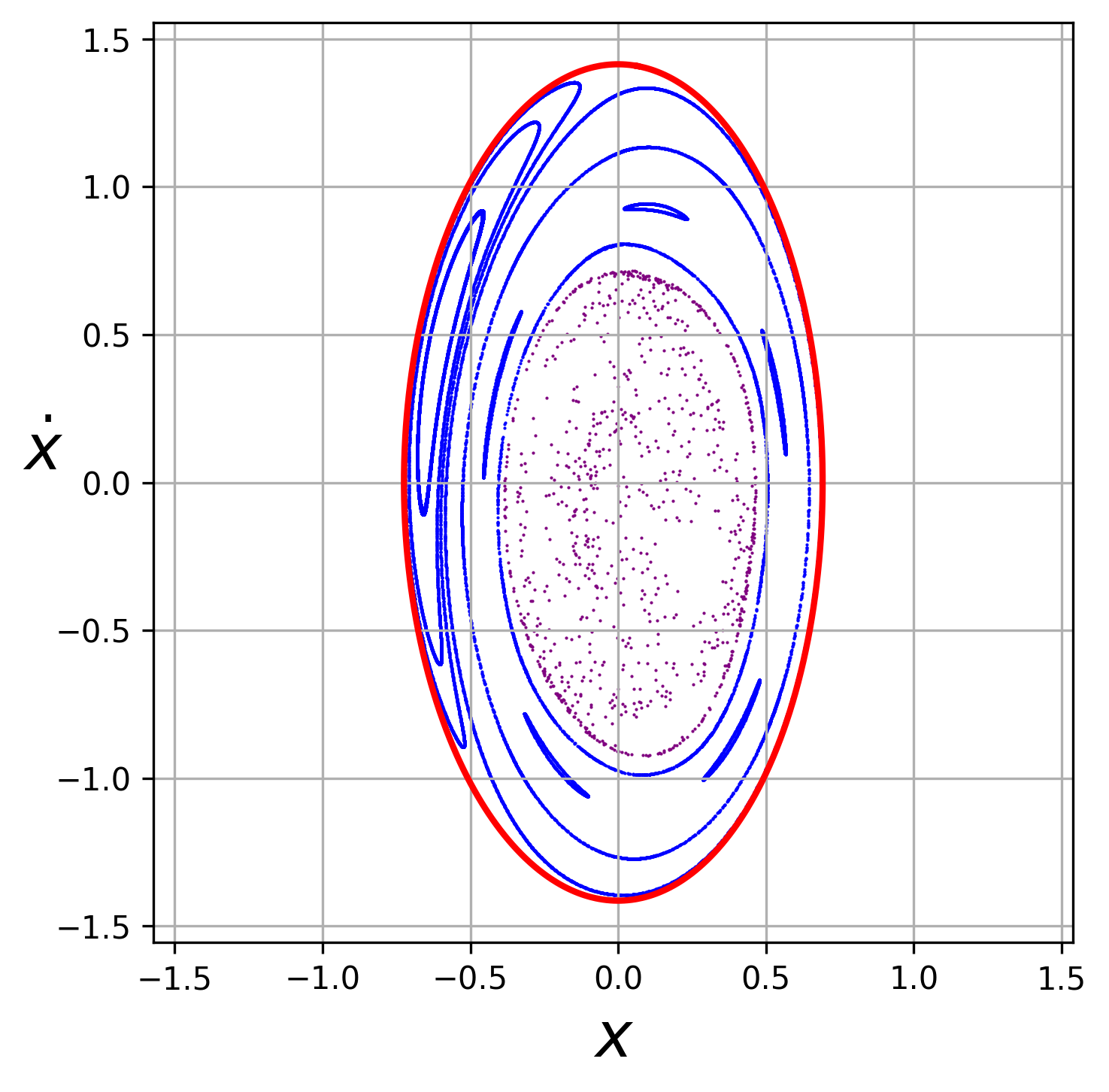}[a]
    \includegraphics[width=0.39\linewidth]{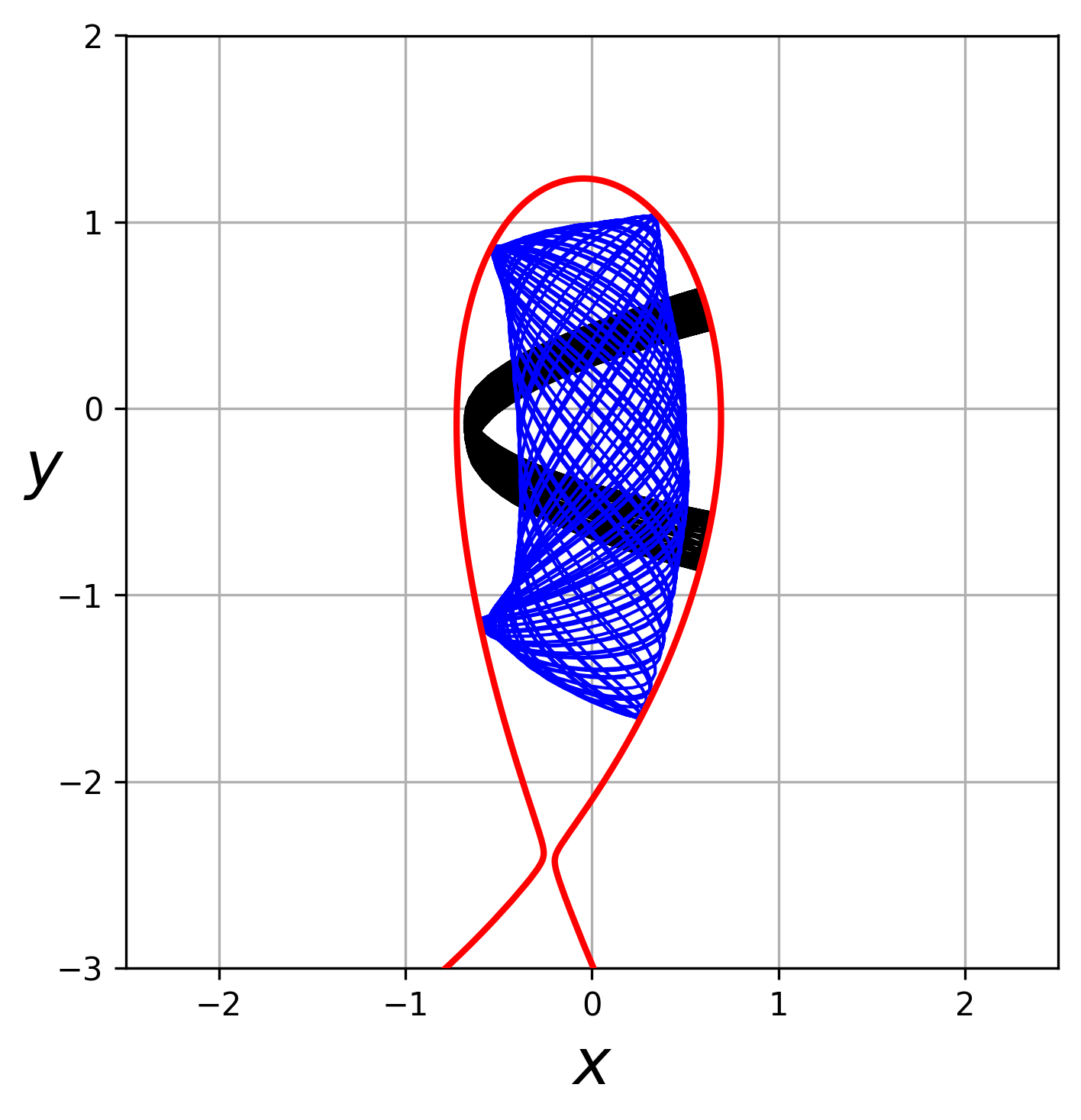}[b]
    \caption{Case with $\epsilon = 0.13$, $\eta=\alpha=\beta=\gamma=1$, $\omega_1=2$, $\omega_2=1$. a) The corresponding Poincar\'{e}  surface of section. b) Two representative Lissajous orbits. Some chaotic orbits escape to $y =-\infty$ through the single opening of the CZV.}
    \label{fig:r5}
\end{figure}

If we increase the perturbation to $\epsilon=0.13$ then we find escapes towards $y^2=\infty$. The invariant curves are given in \figref{r5}a, while the corresponding orbits are shown in \figref{r5}b. The orbits starting on the left  are of Lissajous type, while many orbits starting on the right are chaotic for some time and then most of them escape to $y^2=\infty$. In this case the CZV (red) has only one opening (downwards) in contrast with the case of \figref{r3}a (for $\beta=\gamma=0$) where there are two openings.

The orbits of the fully perturbed case ($\eta=\alpha=\beta=\gamma=1$) are similar to the case ($\alpha=1,\beta=\gamma=0$) for sufficiently small $\beta, \gamma$ and in such cases the orbits are of similar form (ordered Lissajous orbits, chaotic orbits and escaping orbits). For weak perturbation $\epsilon$ the invariant curves and the orbits follow approximately the theoretical results.



\subsection{Special resonance $\omega_1/\omega_2=1/1$ ($\omega_1=\omega_2=1$)}
The case $\omega_1=\omega_2$ has been studied in the past with $H_1=xy^2$ \cite{ContopMouts1966} and in recent years \cite{Contopoulos_2025} with  $H_1=xy^2+ax^3$ ($\eta=1$).
In the latter case we found a formal integral $\phi$ with initial term
\begin{eqnarray}
    \phi_0'=C_0'+\phi_0,
\end{eqnarray}
where 
\begin{align}
C_0'=(2\Phi_{1,0})(2\Phi_{2,0})\cos(2t_0)=x^4-(X^2-x^2)(Y^2-y^2)+XYxy
\end{align}
that does not have secular terms up to order $\phi_2'$ if 
\begin{eqnarray}
    c_1=\frac{45a^2-18a-4}{b(2-a)},\quad c_2=0,\quad c_3=\frac{1-18a}{b^2(2-a)}.
\end{eqnarray}
However, these formulae are not valid if $\alpha=2$. In fact, for $\alpha=2$ and with arbitrary $\omega_1, \omega_2$ this system is integrable \cite{bountis1982integrable,fordy1991henon,lakshmanan2012nonlinear}. 

On the other hand, in the more general case of Eq.~\eqref{pert}
we find secular terms in $\phi_2$ which are proportional to $\sin(2t_0)t, \sin(3t_0)t$ and $\sin(t_0)t$ and it is not possible to eliminate them by any combination of values of $c_1,c_2,c_3$ unless $H_1 = \eta xy^2 + \alpha x^3$ ($\beta=\gamma=0)$ or $H_1 = \beta x^2 y + \gamma y^3$ ($\eta=\alpha=0)$.

The case $\omega_1=\omega_2=1$ is quite special  because it does not follow the rules developed for the resonances $\omega_1/\omega_2=n/m$ in Section 3: a) The secular terms of $\phi_0$ are not of the form $\substack{\sin\\\cos}(mnt_0)t=\substack{\sin\\\cos}(t_0)t$, but $\sin(2t_0)t$, b) the corresponding resonant zero-order integral is not $C_0$ but $C_0'=C_0^2-S_0^2$ (in fact $C_0$ does not give secular terms of the same form as those due to $\phi_0$).

The secular terms generated by $C_0'$ when $\beta=\gamma=0$ are
\begin{align}
C_{2sec}' &= A\left(\frac{15}{2}  \alpha^2 - 3 a \eta - \frac{2 \eta^2}{3} \right)+  B\left( 3\alpha \eta - \frac{1}{6} \eta^2 \right), 
\end{align}
where
\begin{align}
   & A=(2\Phi_{1,0})^2(2\Phi_{2,0}) \sin \left( 2 t_0 \right)t ,\quad
    B=(2\Phi_{1,0})(2\Phi_{2,0})^2 \sin \left( 2 t_0 \right)t.
\end{align}
From $\phi_0$ we find the secular terms
\begin{align}
    \phi_{2sec}=2\eta(-2\eta + \alpha)(2c_1 - c_2)A + 2\eta(-2\eta + \alpha)(c_2 - 2c_3)B,
\end{align}
and thus by setting
\begin{align}
\nonumber c_1 &= \frac{-\left( \frac{15}{2} \alpha^2 - 3\alpha\eta - \frac{2}{3} \eta^2 \right)}{4\eta(-2\eta + \alpha)}, \\\nonumber
c_2 &= 0, \\
c_3 &= \frac{3a\eta - \frac{1}{6} \eta^2}{4\eta(-2\eta + \alpha)}
\end{align}
the secular terms of $\phi_{2}'$ vanish. Similar formulae exist if $\alpha=\eta=0$ and $\beta\gamma\neq 0$. A summary of these results is given in Table 2.

\begin{table}[H]
\centering
\begin{tabular}{|p{3cm}|p{4cm}|p{4cm}|}
\hline
{Case} & {Secular Terms} & {Can They Be Eliminated?} \\
\hline
$C_0$, with arbitrary $ \eta\alpha\beta\gamma\neq 0$.& 
Secular terms appear in $\phi_1$. & 
No, they cannot be eliminated. \\
\hline
$\eta\alpha\neq 0$, $\beta=\gamma=0$ & 
Secular terms of degree 6. & 
Yes, the secular terms are eliminated by appropriate $c_1,c_2,c_3$ in $\phi_0'$.
 \\
\hline
$\beta\gamma\neq 0$ and $\eta=\alpha=0$ & 
Secular terms of degree 6. & 
Yes, the secular terms are eliminated by appropriate $c_1,c_2,c_3$ in $\phi_0'$. \\
\hline
$S_0$ with arbitrary parameters & 
Secular terms 
appear in $S_2$.& 
No, they cannot be eliminated. \\
\hline
\end{tabular}
\caption{Presence and eliminability of secular terms in the case \(\omega_1/\omega_2 = 1/1\).}
\end{table}

\begin{figure}[H]
    \centering
    \includegraphics[width=0.4\linewidth]{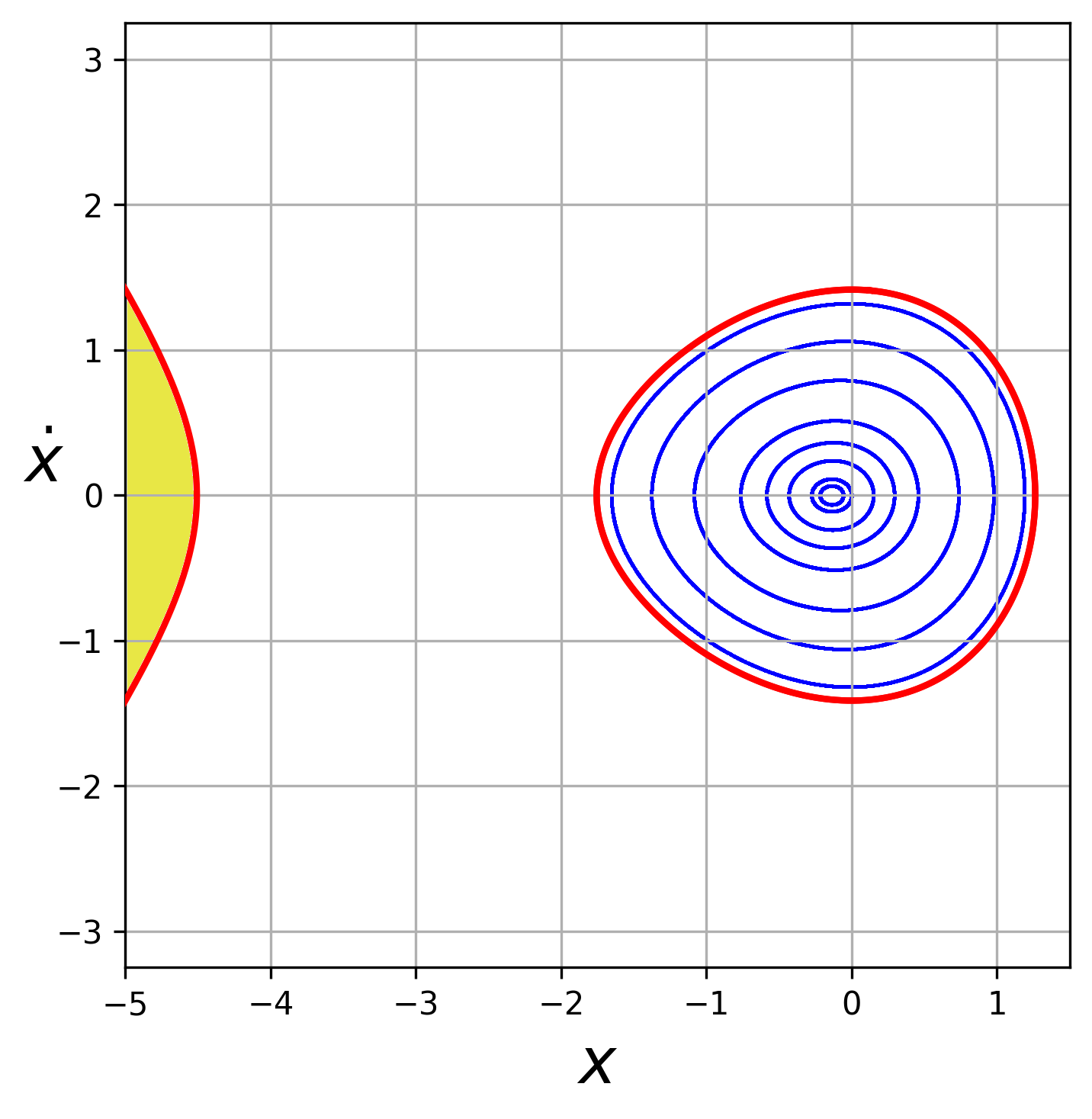}
    \includegraphics[width=0.415\linewidth]{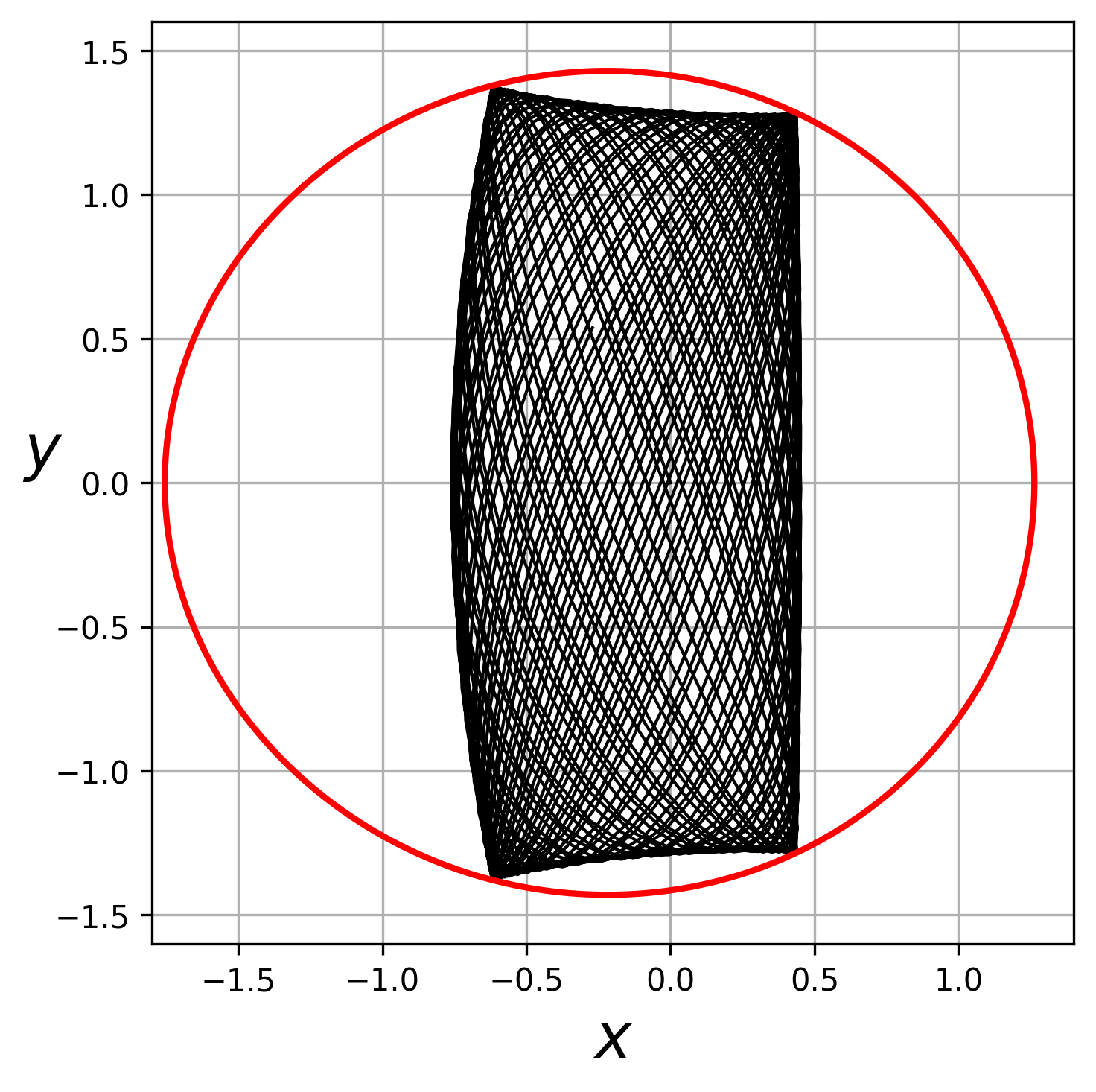}
    \caption{Case with $\epsilon = 0.1$, $\eta=\alpha= 1$, $\beta=\gamma=0$, $\omega_1=\omega_2=1$. (a) Theoretical invariant curves (red) and exact (numerical) (blue). The yellow region on the left corresponds to escaping orbits. (b) Almost all the orbits are of Lissajous form. The red curve is the CZV.}
    \label{fig:r6}
\end{figure}

The surface of section for $\epsilon=0.1$, $\eta=\alpha=1,\beta=\gamma=0, \omega_1=
\omega_2=1$ is given in \figref{r6}a. This figure contains exact (numerical) invariant curves (blue) and  theoretical invariant curves (up to order $\epsilon^2$) (red). The agreement between the corresponding invariant curves is very good, while the corresponding orbits are Lissajous figures (\figref{r6}b).

\begin{figure}[H]
    \centering
    \includegraphics[width=0.4\linewidth]{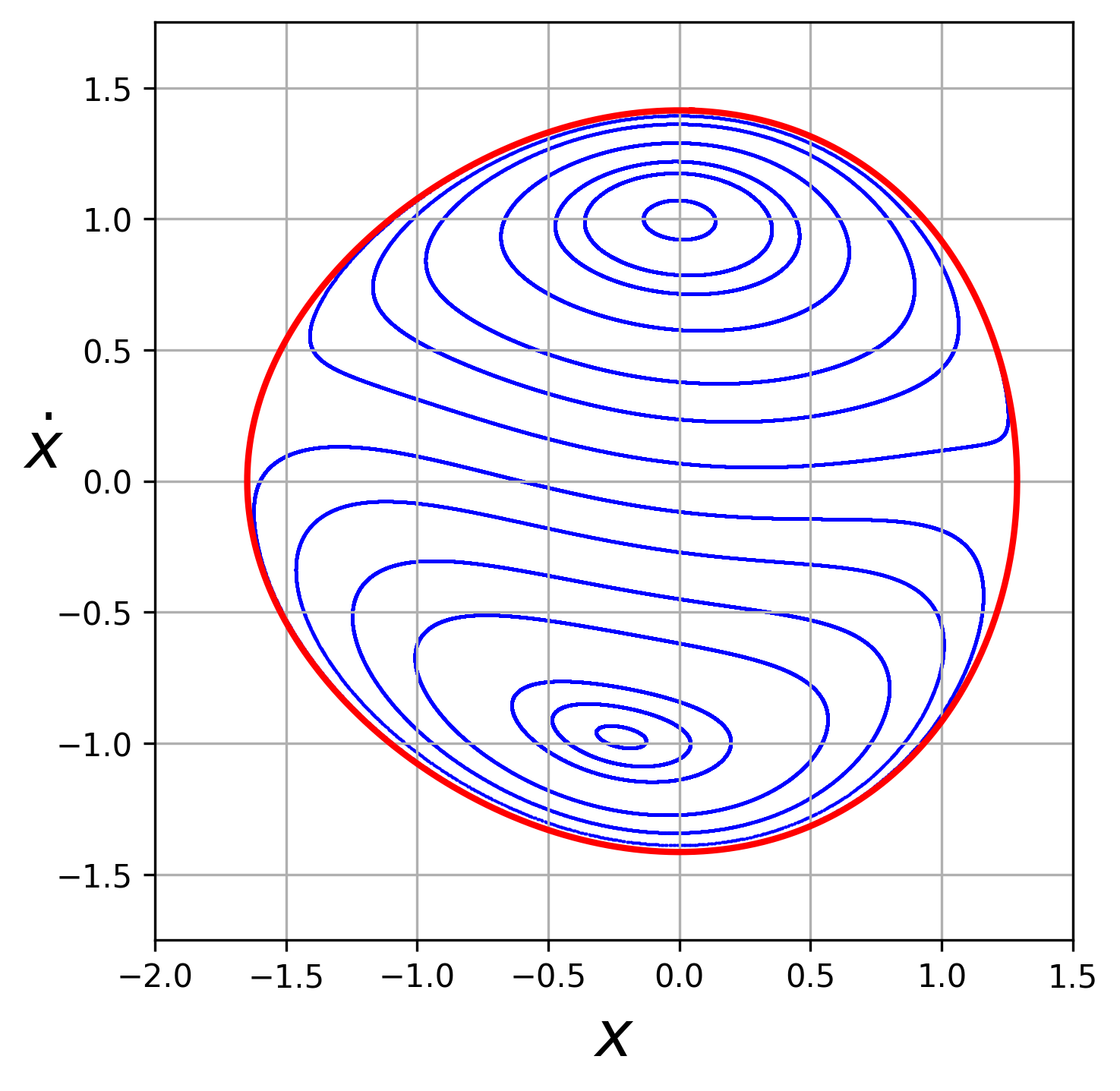}[a]
    \includegraphics[width=0.4\linewidth]{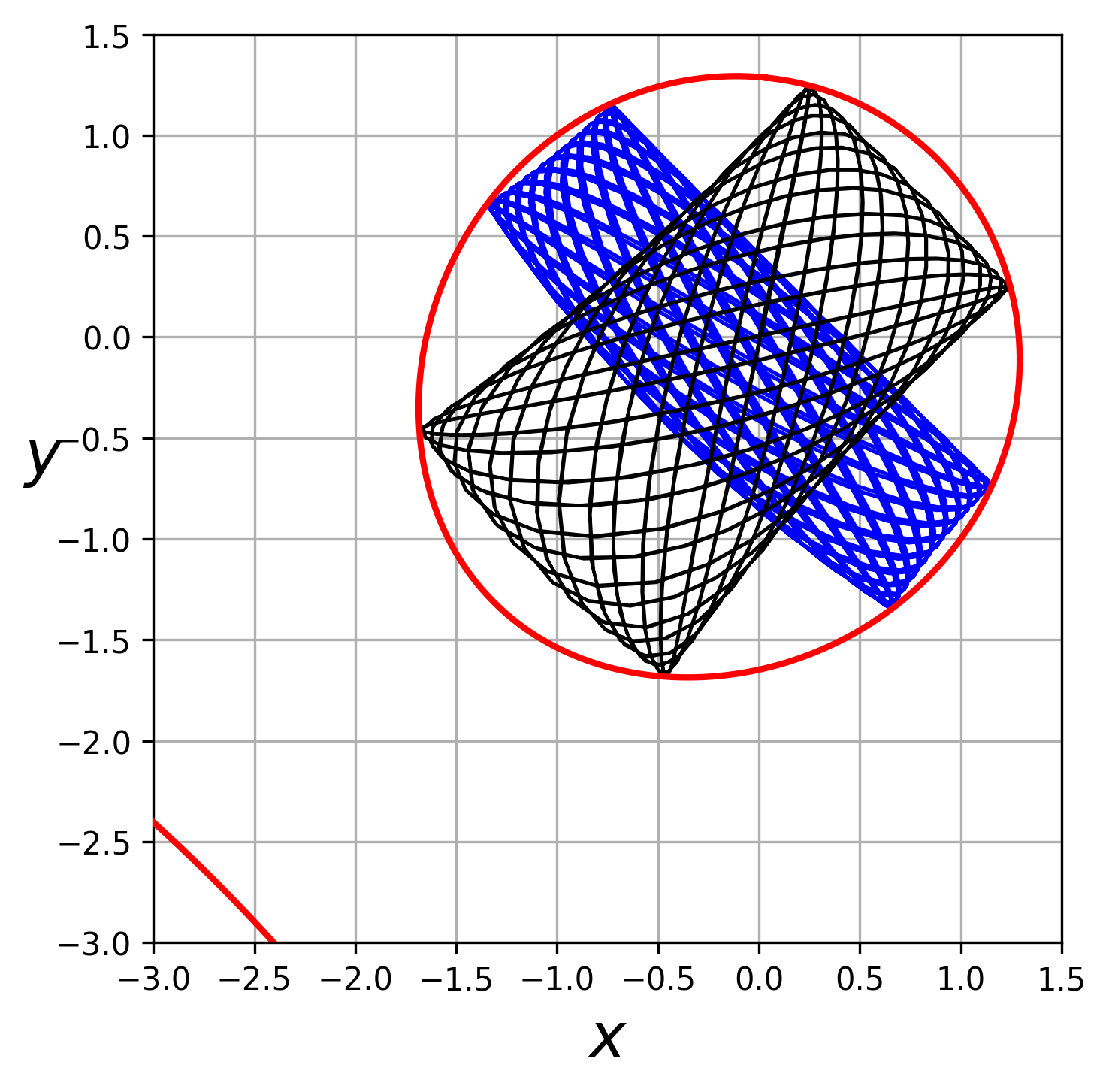}[b]
    \caption{Case with $\epsilon = 0.08$, $\eta=\alpha= \beta=\gamma=1$, $\omega_1=\omega_2=1$. a) The Poincar\'e surface of section $y=0$. b) Two representative Lissajous orbits inside the CZV (red curve).}
    \label{fig:r7n}
\end{figure}

Then in \figref{r7n}a we give the surface of section for $\epsilon=0.1,\alpha=\beta=\gamma=1,\omega_1=\omega_2=1$ and in \figref{r7n}b we give two Lissajous orbits. Here there are no escapes.  However, there is no formal integral to give the corresponding theoretical invariant curves. In fact, if we use $\phi_0$ as zero-order integral then  we find secular terms with $\cos(3t_0)t, \sin(2t_0)t$ and $\sin(t_0)t$ that cannot be eliminated by any choice of the constants $c_1,c_2,c_3$.

\begin{figure}[H]
    \centering
    \includegraphics[width=0.385\linewidth]{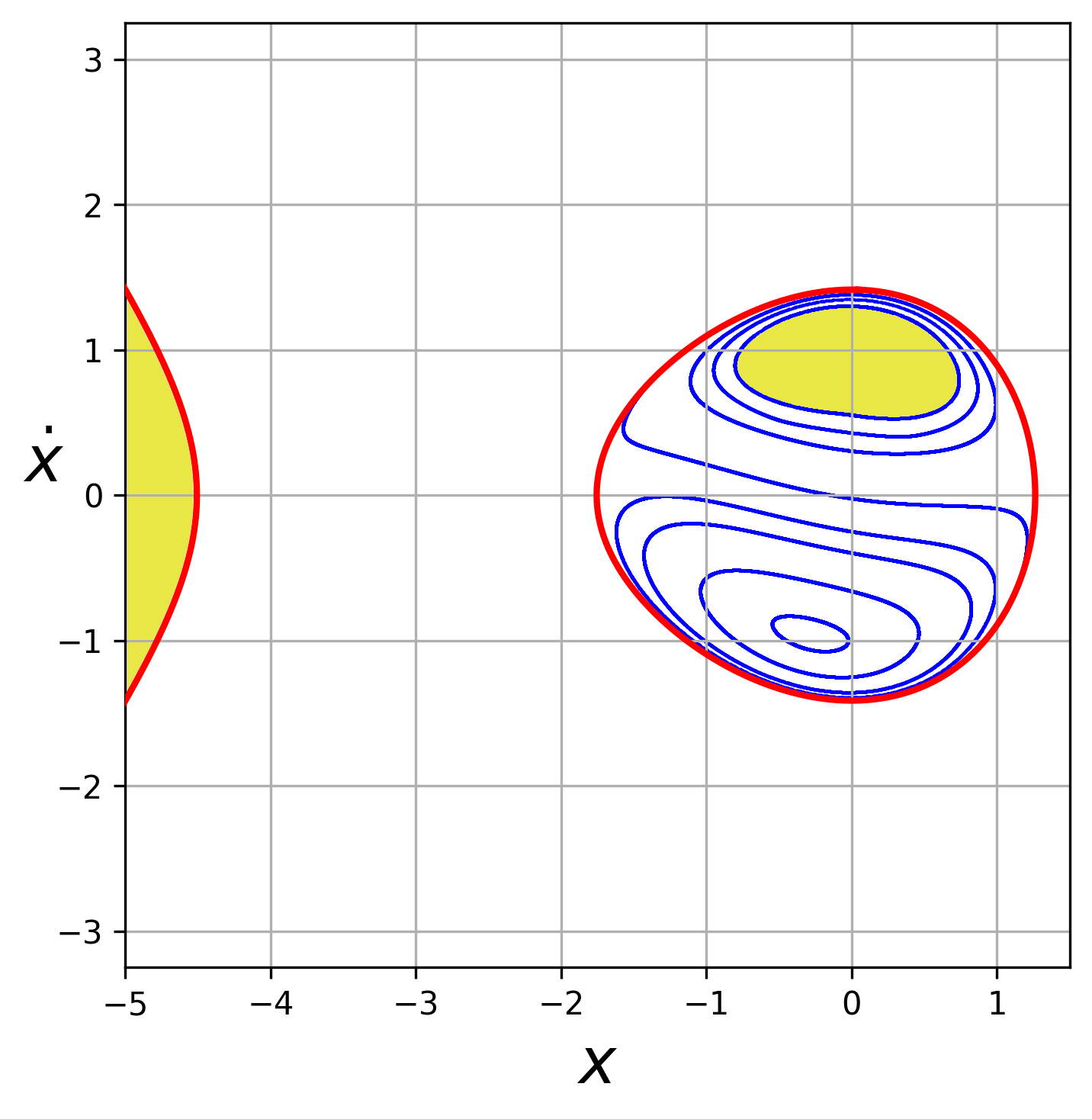}[a]
    \includegraphics[width=0.4\linewidth]{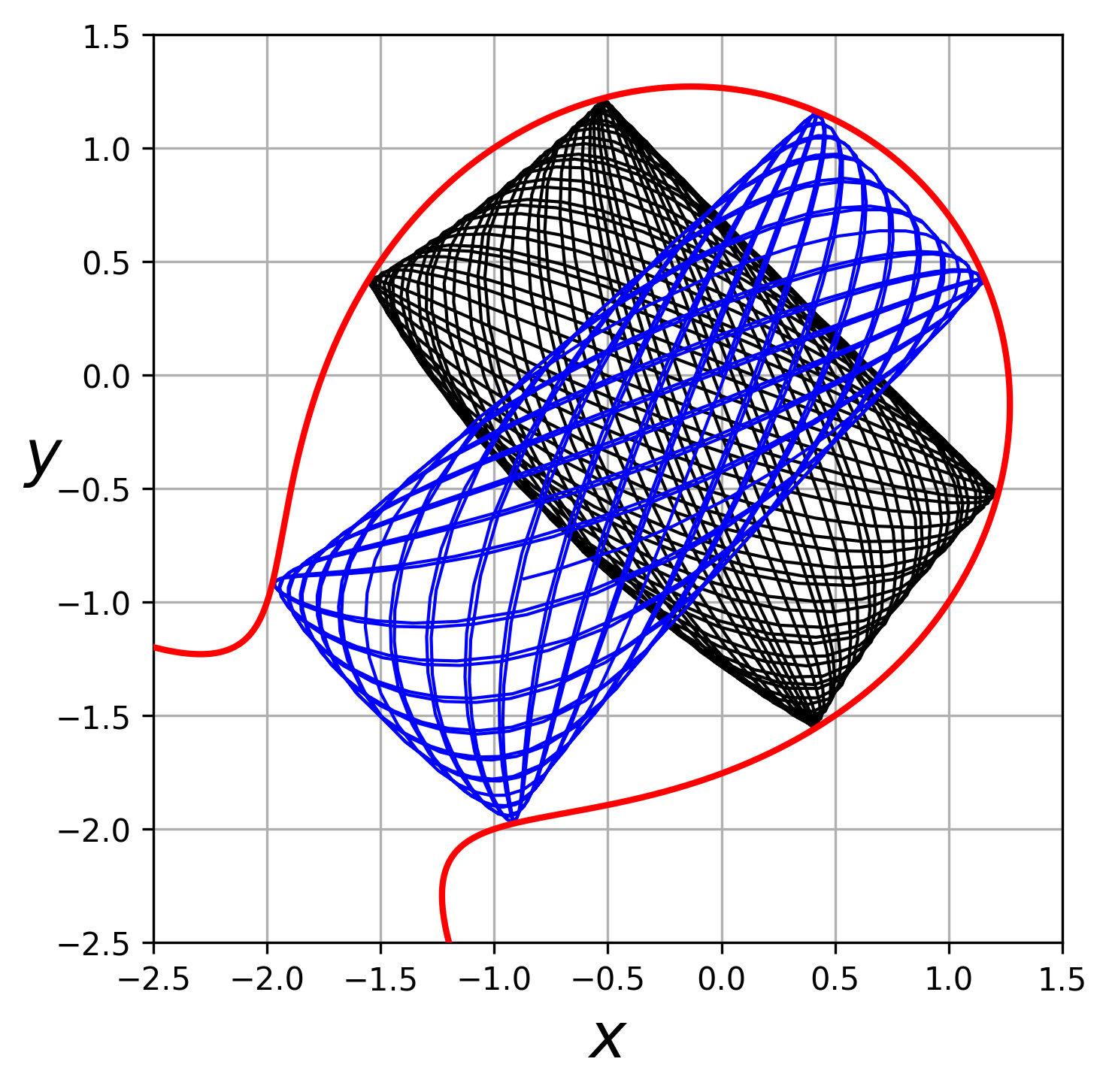}[b]
    \caption{Case with $\epsilon = 0.1$, $\eta=\alpha=\beta=\gamma=1$, $\omega_1=\omega_2=1$. (a) The corresponding Poincar\'{e} surface of section. Numerical (exact) invariant curves and a region of escaping orbits towards $y^2 = \infty$ (yellow). (b) Two representative Lissajous orbits. Some orbits escape through the opening of the CZV down left.}
    \label{fig:r7}
\end{figure}

In \figref{r7}a,b we give the invariant curves and the orbits for $\epsilon=0.1$ and $\eta=\alpha=\beta=\gamma=1$ ($\omega_1=\omega_2=1)$. We see some invariant curves and regions of escapes (yellow). We have both Lissajous orbits and escaping orbits, while  the CZV has only one opening.


In order to approximate these invariant curves and orbits we calculate the surface of section and the orbits for $\epsilon=0.1, \alpha=\beta=\gamma=1,\omega_1=5\sqrt{2}/7=1.010\dots,\omega_2=1$ by using the integrals of the non resonant case of Section 2.

\begin{figure}[H]
    \centering
    \includegraphics[width=0.4\linewidth]{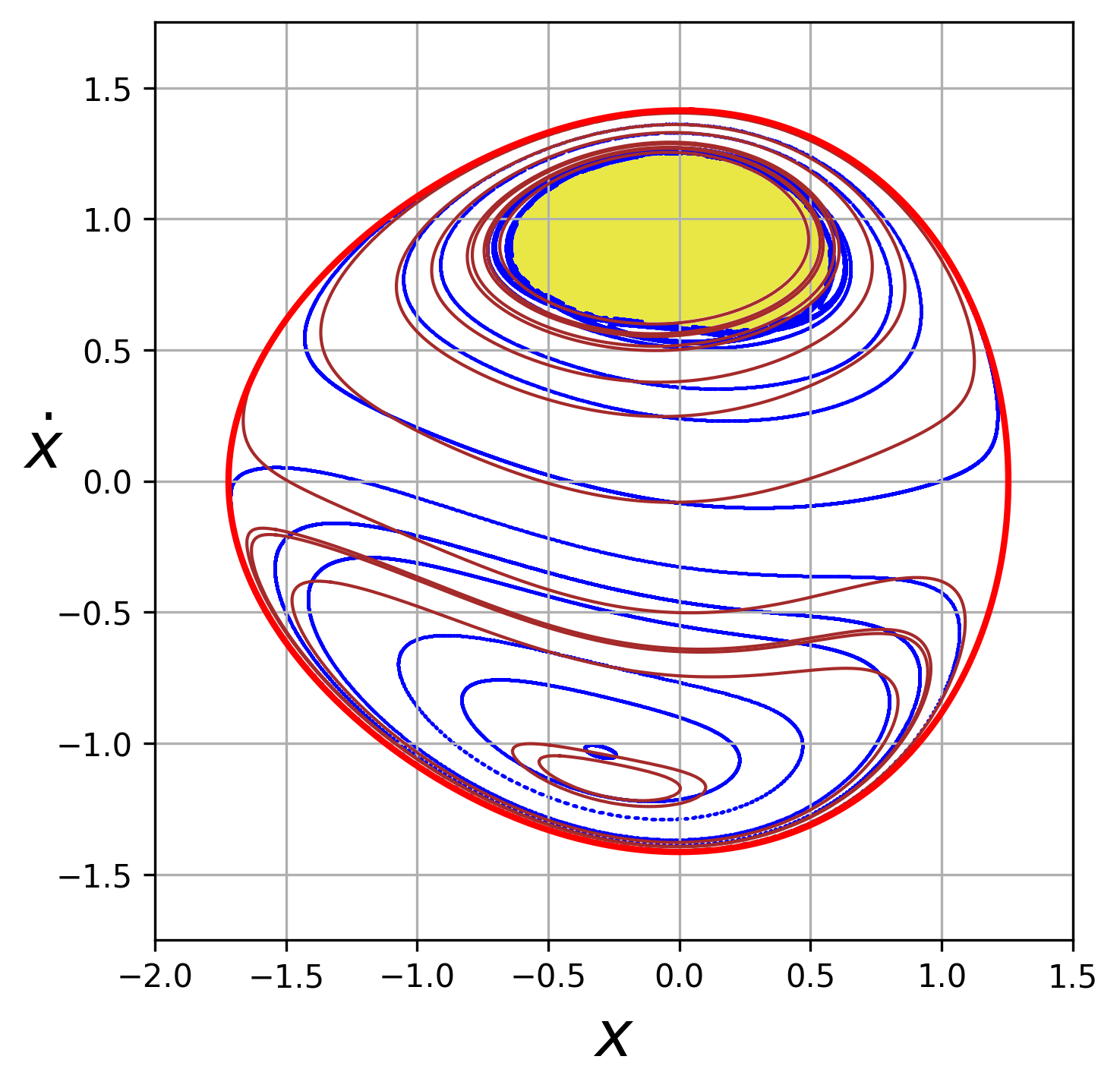}[a]
    \includegraphics[width=0.398\linewidth]{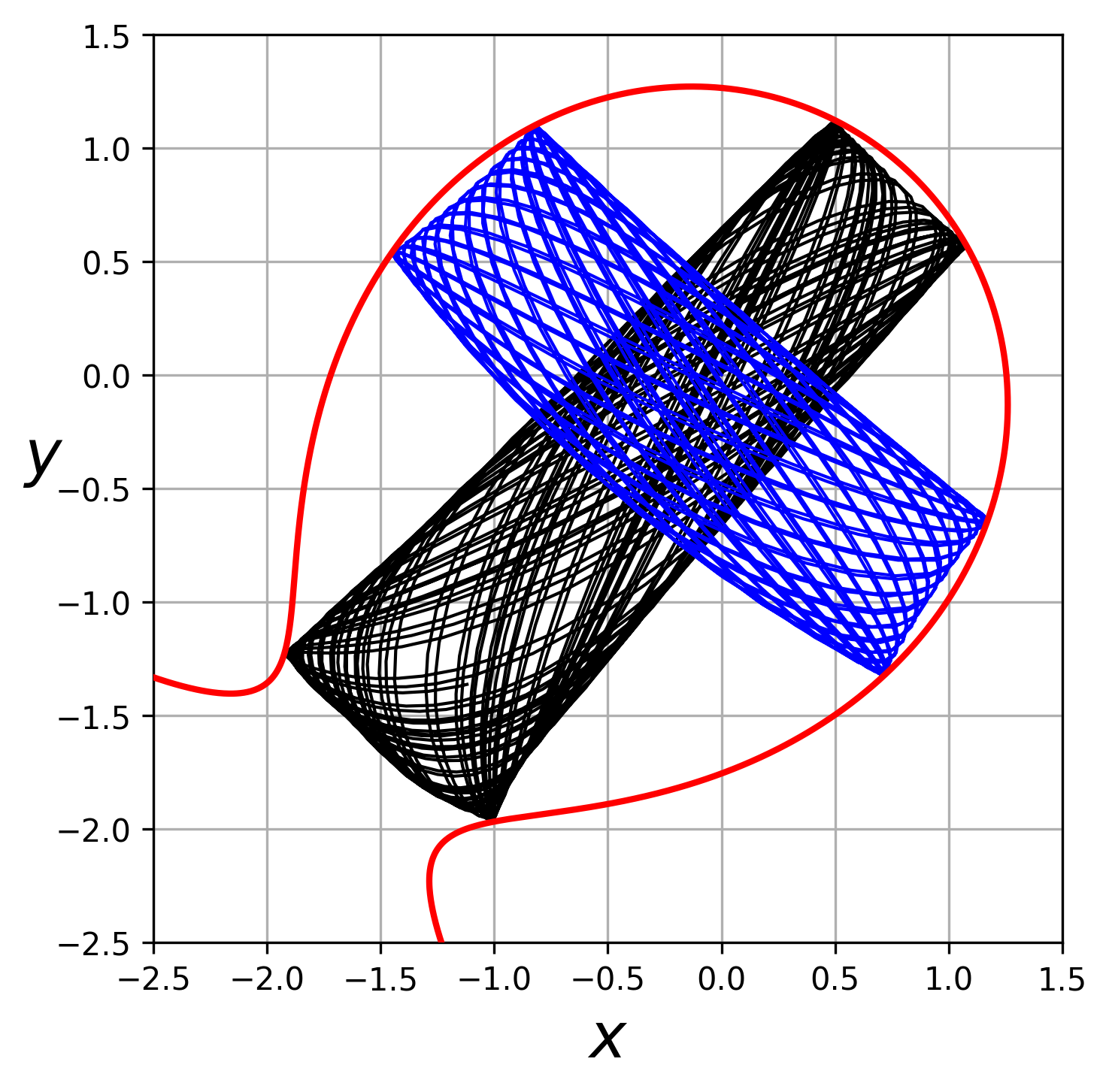}[b]
    \caption{Case with $\epsilon = 0.1$, $\eta=\alpha=\beta=\gamma=1$, $\omega_1=5\sqrt{2}/7$, $\omega_2=1$. (a) The corresponding Poincar\'{e} surface of section. Exact invariant curves (blue) and theoretical (red). (b) Two typical orbits. Some orbits escape through the down left opening of the CZV.}
    \label{fig:r8}
\end{figure}

Then we find the invariant curves of \figref{r8}a, exact (blue) and theoretical (red) and a region of escapes (yellow). The theoretical curves are close to the exact ones and they are similar to the exact invariant curves of the case $\omega_1=\omega_2=1$ (\figref{r7}a).

The corresponding orbits are given in \figref{r8}b and are very similar to the orbits of \figref{r7}b ($\omega_1=\omega_2=1)$. The CZV has again one opening downwards and the orbits passing through it escape to $y=-\infty$.

\begin{figure}[H]
    \centering
    \includegraphics[width=0.44\linewidth]{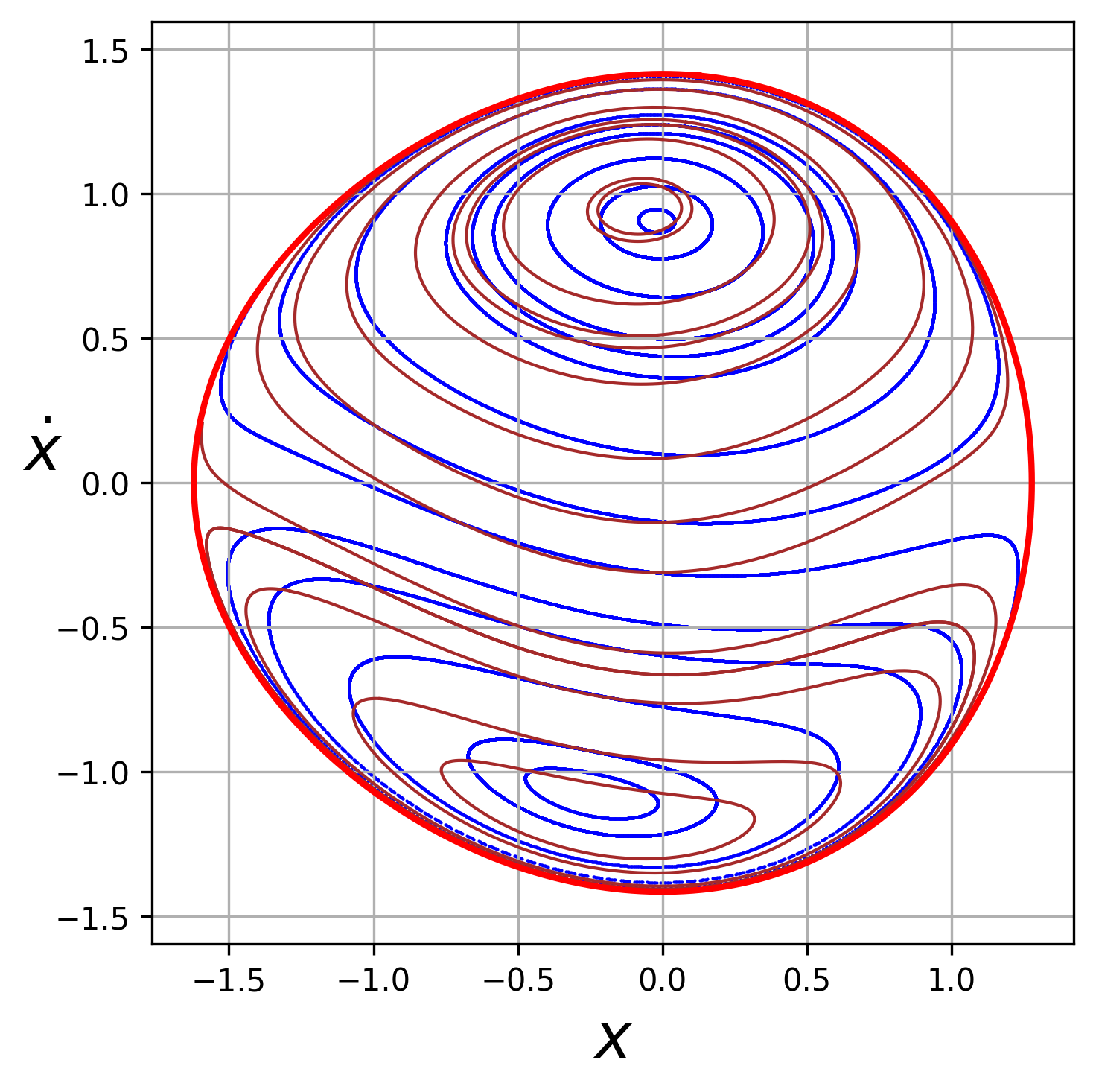}[a]
     \includegraphics[width=0.45\linewidth]{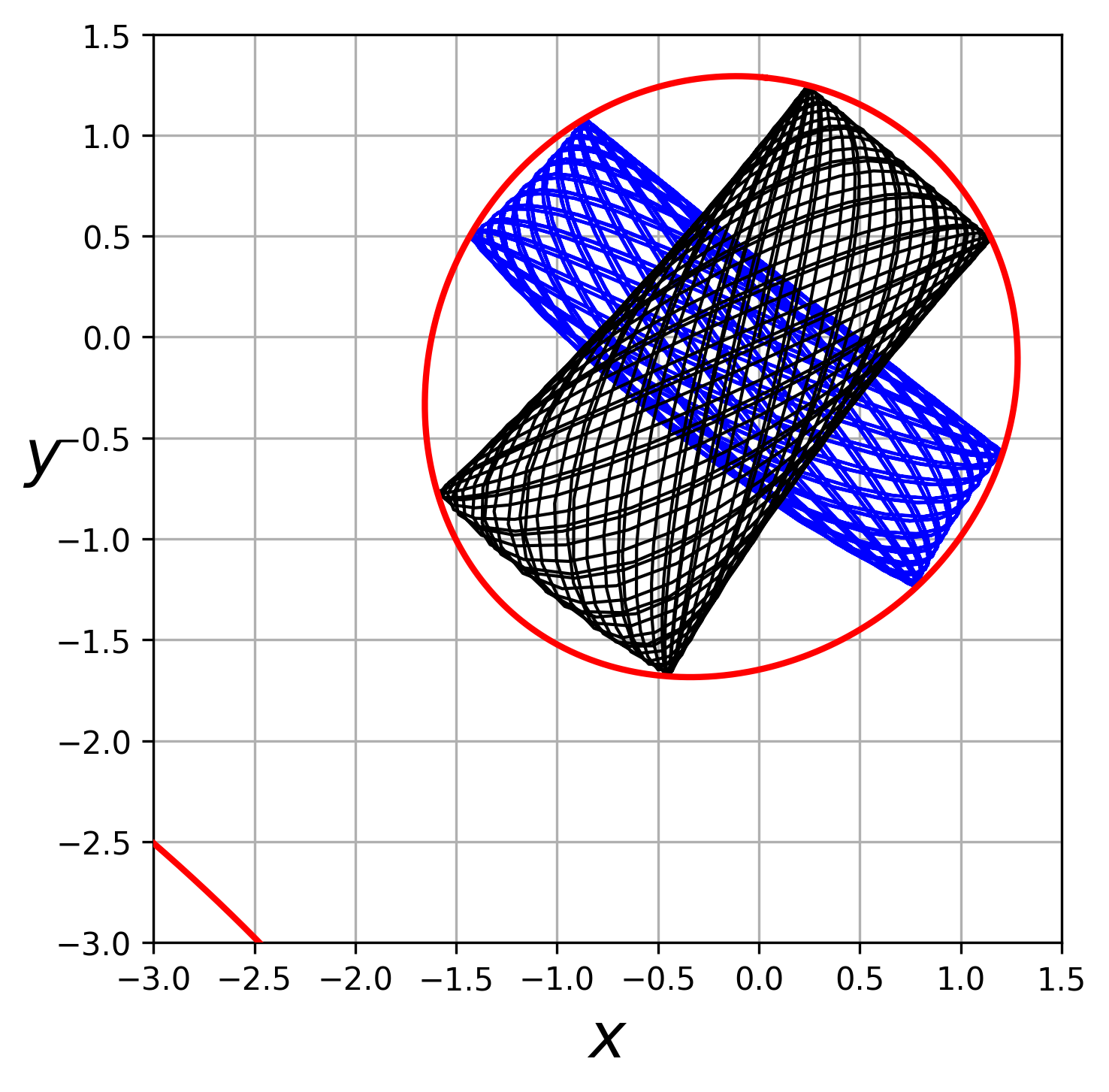}[b]
    \caption{Case with $\epsilon = 0.08$, $\eta=\alpha=\beta=\gamma=1$, $\omega_1=5\sqrt{2}/7$, $\omega_2=1$. (a) The corresponding Poincar\'{e} surface of section. Theoretical (red) and exact (blue) invariant curves similar to those of \figref{r7n}a ($\omega_1=\omega_2=1$). b) Most orbits are of Lissajous form, as in \figref{r7n}b.}
    \label{fig:r9}
\end{figure}

If we take $\epsilon = 0.08$ (and keep the other constants the same) then there are no escapes (\figref{r9}a). The theoretical invariant curves (red) are closer to the exact ones (blue) and all the orbits are approximately Lissajous \figref{r9}b. We conclude that the particular case $\omega_1=\omega_2=1$ is well approximated by the non-resonant case $\omega_1=5\sqrt{2}/7=1.010\dots, \omega_2=1$.

\section{Conclusions}

In the present paper we studied the limits of applicability of the theory of the third integral of Galactic Dynamics by considering a potential $H=H_0+\epsilon H_1$ where $H_0$ represents two harmonic oscillators with frequencies $\omega_1, \omega_2$ and a $H_1$ is a general  perturbation of third degree $H_1=\eta xy^2+\alpha x^3 +\beta x^2y +\gamma y^3$.

In \cite{Contopoulos1966a} we had studied perturbations of the form $H_1 = x^m y^n$, extending the analysis to high orders. More recently, we considered perturbations of the form $ H_1 = x y^2 + \alpha x^3$, which are symmetric with respect to the $ y = 0$ axis, under the condition $ \omega_1 = \omega_2 = 1$.

Here we  studied the more general case where  $\eta, \alpha, \beta, \gamma$ are all nonzero. We derived the general form of the third integral of motion for the non-resonant case, where $\omega_1 / \omega_2$  is irrational. In this case, the integral contains terms with denominators of the form $(m^2 \omega_1^2 - n^2 \omega_2^2)$. However, in the resonant case where $\omega_1 = m$ and $\omega_2 = n$ , these denominators vanish, rendering this form of the integral invalid. Nevertheless, an additional zero-order integral of motion $\substack{S_0\\C_0}=(2\Phi_{1,0})^{m/2}(2\Phi_{2,0})^{n/2}\substack{\sin\\\cos}(mt_0)$ exists in this case.

Then if we construct the successive terms of the integrals
$\Phi=\Phi_{0}+\epsilon\Phi_1+\dots\epsilon^{m+n-2}\Phi_{m+n-2}$,
where $\Phi_0$ is either $\Phi_{1,0}=\frac{1}{2}(\omega_1^2x^2+X^2)$ or $\Phi_{2,0}=\frac{1}{2}(\omega_2^2y^2+Y^2)$, and $\substack{S\\C}=\substack{S_0\\C_0}+\epsilon\substack{S_1\\C_1}+\epsilon^2\substack{S_2\\C_2},$
we find secular terms containing the factor $\substack{\sin\\\cos}(mnt_0)$ in $\Phi_{m+n-2}$ and in $\substack{S_2\\C_2}$.
The term $\Phi_{m+n-2}$ is of degree $m+n$ in $x,y,X,Y$ while $S_2$ (or $C_2$) is of degree $m+n+2$. But if we use as zero-order integral $\phi_0=c_1(2\Phi_{1,0})^2+c_2(2\Phi_{1,0})(2\Phi_{2,0})+c_3(2\Phi_{2,0})^2$
then we find secular terms of degree $m+n+2$ in $x,y,X,Y$. As regards  the powers of $\epsilon$, the secular terms of $\substack{S_2\\C_2}$ acquire  the same form as $\epsilon^{m+n-2}\phi_{m+n-2}$ if we multiply $S_0$ (or $C_0$) by $\epsilon^{m+n-4}$. Then by adding the series produced by $\phi_0$ and $\epsilon^{m+n-4}\substack{S_0\\C_0}$ and taking appropriate values of $c_1,c_2,c_3$ we eliminate the secular terms of degree $m+n+2$ in $x,y,X,Y$.

This is the general method used for eliminating secular terms. In our present paper we considered in particular the representative cases $4/1, 5/2,3/2$ and $4/3$. However, there are also some special cases that need a particular treatment such as $3/1,2/1$ and $1/1$.

In the case $\omega_1=3, \omega_2=1$ we found a formal integral with zero-order term $\Phi_0+C_0$ and $\beta\gamma\neq 0$. But when $\beta=\gamma=0$ this form of the  integral diverges. However, in this particular case the formal integral $\Phi=\Phi_{1,0}+\epsilon\Phi_{1,1}+\dots$ does not have secular terms because the denominator $(9\omega_1^2-\omega_2^2)$ of the irrational expansion is eliminated by a similar factor in the numerator of $\Phi_{1,2}$. Thus, in this particular case, the third integral is $\Phi$ itself.

In the case $\omega_1=2,\omega_2=1$ the resonant integral has a lowest order term $S_0$.

Finally, in the case $\omega_1=\omega_2=1$ we can construct an integral if $\beta=\gamma=0$ (this was done in our previous paper \cite{Contopoulos_2025}) but we cannot eliminate all the secular terms if $\beta\gamma\neq0$. In fact in this case there are secular terms containing $\sin(2t_0)t$ but also terms with $\sin(3t_0)t$ and $\sin(t_0)t$. In this case we could use as an approximation the non-resonant integral with $\omega_1=5\sqrt{2}/7=1.010...$

We did not find  another case with more than one type of secular terms.
Therefore, it seems that the only case in which we cannot construct a formal integral of motion is when $\omega_1=\omega_2$ and $\eta\alpha\beta\gamma\neq0$.

In all cases (both resonant and non-resonant) the main reason for 
the non-applicability of the formal integral is the appearance of chaos when the perturbation $\epsilon$ is strong. Chaos is generated by the overlapping of resonances \cite{Rosenbluth1966,Contopoulos1966c}. Namely, the asymptotic curves from various unstable periodic orbits intersect at an infinity of homoclinic and heteroclinic orbits. An example of such intersection was given in \cite{Contopoulos_2025} for the original H\'{e}non-Heiles potential.

In the present paper we studied the onset of chaos in the last two cases considered above, i.e. $\omega_1/\omega_2=2/1$ and $1/1$. We found the increase of chaos when $\epsilon$ increases and the differences between the cases of symmetric perturbation (when $\beta=\gamma=0$) and asymmetric perturbations when $\beta\gamma\neq0$. The invariant curves generated by the points of the orbits when $y=0$ are approximated by the theoretical curves given by the formal integral when the orbits are regular (generalized Lissajous figures) but the points are scattered when the orbits are chaotic.

Furthermore the orbits may escape to infinity when the curves of zero velocity are open. In such cases most chaotic orbits may escape to infinity although for some time they may look approximately as ordered.

In particular, we found that in the symmetric cases ($\beta=\gamma=0$) the escapes start, in general, through two symmetric openings of the CZV, above and below the $y=0$ axis, while in the asymmetric cases ($\beta\gamma\neq 0$) we found a single opening that allows escapes through it.

The onset of chaos for other values of $\omega_1/\omega_2$ has the same qualitative characteristics as in the cases $\omega_1/\omega_2=2/1$ and $1/1$ considered above.

A remarkable theoretical result that was found by Moser \cite{Moser1956,Moser1958} and Giorgilli \cite{Giorgilli2001} is that in the chaotic region, close to unstable periodic orbits, the formal integral becomes convergent. Thus, the successive points of the chaotic orbits can be followed analytically. This topic has been studied in several papers up to now \cite{ritter1987analytical,de1997extended,vieira1996study,EfthContKats2014} and we will consider it further in the future.

\section*{Appendix: A comment on the calculation of the resonant formal integrals}
The construction of a resonant formal integral of motion depends heavily on our ability to pass from functions of $x,y,X,Y$ to  functions of $t,t_0$. 
In our algorithm we first  use the forward transformations 
$F_1 := x = \frac{\sqrt{2 \Phi_{1,0}}}{\omega_1} \sin\left(\omega_1(t - t_0) \right), 
F_2 := y = \frac{\sqrt{2 \Phi_{2,0}}}{\omega_2} \sin\left(\omega_2 t \right), 
F_3 := X = \sqrt{2 \Phi_{1,0}} \cos\left(\omega_1 (t - t_0) \right), 
F_4 := Y = \sqrt{2 \Phi_{2,0}} \cos\left(\omega_2 t \right).$
in order to pass from the spatial dependence to time dependence and after the integration the backward transformations
$
B_1 := \sin(\theta) = \frac{x \omega_1}{\sqrt{2 \Phi_{1,0}}}, 
B_2 := \cos(\theta) = \frac{X}{\sqrt{2 \Phi_{1,0}}}, 
B_3 := \sin(\omega_2 t) = \frac{y \omega_2}{\sqrt{2 \Phi_{2,0}}}, 
B_4 := \cos(\omega_2 t) = \frac{Y}{\sqrt{2 \Phi_{2,0}}}.$
where $\theta=\omega_1(t-t_0)$. 

By use of a computer algebra system (we work with Maple) we need in all cases to expand the resulting expression after the integration with respect to $t$,  in order to apply $B_1,B_2,B_3,B_4$. At this step there is a technical pitfall: Most computer algebra systems when asked to expand expressions with trigonometric terms of the form $\cos(kt),\sin(kt)$ etc. apply recursively the formula connecting these terms with Chebyshev polynomials of first kind
\begin{align}
    \cos(mt) = T_m(\cos t),
\end{align}
where
\begin{align}
T_m(x) = \frac{m}{2} \sum_{k=0}^{\left\lfloor \frac{m}{2} \right\rfloor} 
\frac{(-1)^k (m-k-1)!}{k!(m - 2k)!} (2x)^{m - 2k}, \quad m \geq 1.
\end{align}
By doing so we always find powers of $\cos(t)$ and $\sin(t)$. This does not allow to a straightforward application of $B_3$ and $B_4$. Thus, it is necessary to introduce $u=\omega_2t\to t=u/\omega_2$ in order to prevent the software from expanding $\omega_2t$ when $\omega_2>1$.

\section*{Acknowledgements}
This work was funded by the Sectoral Development Program ($O\Pi\Sigma 5223471$) of the Ministry of Education, Religious Affairs and Sports, through the National Development Program (NDP) 2021-25. It was conducted as part of project 200/1022, “Non linear phenomena in Galactic Disks”, supported by the Research Committee of the Academy of Athens.

\bibliographystyle{iopart-num}
\bibliography{bibliography}

\end{document}